\documentclass[twocolumn,superscriptaddress,floatfix,
longbibliography,aps,pra]{revtex4-2}

\usepackage[utf8]{inputenc}
\usepackage[T1]{polski}
\usepackage[T1]{fontenc}
\usepackage[english]{babel}

\usepackage{graphicx}
\usepackage{bm}
\usepackage{xcolor}
\usepackage{epstopdf}
\usepackage{amsmath}
\usepackage{amssymb}
\usepackage{epstopdf}
\usepackage[normalem]{ulem} 
\usepackage{enumerate}

\usepackage[urlcolor=blue,colorlinks=true,citecolor=blue,linkcolor=blue,pdfstartview={FitH},bookmarks=false]{hyperref}

\newcommand{\Tr}{\text{Tr}}

\newcommand{\Imag}{\text{Im}}
\newcommand{\e}{\textrm{e}}
\newcommand{\im}{\textrm{i}}
\newcommand{\brho}{\bm{\rho}}
\newcommand{\bsig}{\bm{\sigma}}
\newcommand{\btau}{\bm{\tau}}

\graphicspath{{fig/}{./fig/}{.}}

\DeclareMathOperator{\pf}{Pf}
\DeclareMathOperator{\sgn}{sgn}

\begin{document}

\title{Majorana bound states in a superconducting Rashba nanowire \\
in the presence of antiferromagnetic order}

\author{Aksel Kobia\l{}ka}
\email[e-mail: ]{akob@kft.umcs.lublin.pl}
\affiliation{Institute of Physics, Maria Curie-Sk\l{}odowska University, 
Plac Marii Sk\l{}odowskiej-Curie 1, PL-20031 Lublin, Poland}

\author{Nicholas Sedlmayr}
\email[e-mail: ]{sedlmayr@umcs.pl}
\affiliation{Institute of Physics, Maria Curie-Sk\l{}odowska University, 
Plac Marii Sk\l{}odowskiej-Curie 1, PL-20031 Lublin, Poland}

\author{Andrzej Ptok}
\email[e-mail: ]{aptok@mmj.pl}
\affiliation{Institute of Nuclear Physics, Polish Academy of Sciences, 
ul. W. E. Radzikowskiego 152, PL-31342 Krak\'{o}w, Poland}

\date{\today}

\begin{abstract}
Theoretical studies have shown that Majorana bound states can be induced at the ends of a one dimensional wire, a phenomenon possible due to the interplay between s-wave superconductivity, spin--orbit coupling, and an external magnetic field. 
These states have been observed in superconductor--semiconductor hybrid nanostructures in the presence of a Zeeman field, and in the limit of a low density of particles. 
In this paper, we demonstrate and discuss the possibility of the emergence of Majorana bound states in a superconducting Rashba nanowire deposited on an antiferromagnetically ordered surface.
We calculate the relevant topological invariant in several complementary ways. 
Studying the topological phase diagram reveals two branches of the non trivial topological phase---a main branch, which is typical for Rashba nanowires, and an additional branch emerging due to the antiferromagnetic order.
In the case of the additional topological branch, Majorana bound states can also exist close to half-filling, obviating the need for either doping or gating the nanowire to reach the low density regime.
Moreover, we show the emergence of the Majorana bound states in the absence of the external magnetic field, which is possible due to the antiferromagnetic order.
We also discuss the properties of the bound states in the context of real space localization and the spectral function of the system.
This allows one to perceive the band inversion within the spin and sublattice subspaces in the additional branch, contrary to the main branch, where the only band inversion reported in previous studies exists in the spin subspace.
Finally, we demonstrate how these topological phases can be confirmed experimentally in transport measurements.

\end{abstract}

\maketitle

\section{Introduction}
\label{intro}

The possibility for topologically protected localized zero energy states to form in a superconducting nanowire was first proposed in a seminal paper by Kitaev~\cite{kitaev.01}, and opened a period of intense study of these Majorana bound states (MBS)~\cite{aguado.17,lutchyn.bakkers.18,pawlak.hoffman.19}. 
The states are of particular interest because they are non-Abelian anyons, and thus potentially of interest for topological quantum computing~\cite{nayak.simon.08}. 
In the last decade, potential signatures of MBS have been detected in low--dimensional structures, e.g.~semiconducting--superconducting hybrid nanostructures~\cite{deng.yu.12,mourik.zuo.12,das.ronen.12,finck.vanharlingen.13,nichele.drachmann.17,gul.zhang.18,deng.vaitiekenas.16,deng.vaitiekenas.18} and chains of magnetic atoms deposited on a superconducting surface~\cite{nadjperge.drozdov.14,pawlak.kisiel.16,feldman.randeria.16,ruby.heinrich.17,jeon.xie.17,kim.palaciomorales.18}. 
In the first case it is the interplay between intrinsic spin--orbit coupling (SOC), proximity induced superconductivity, and an external magnetic field, which leads to the emergence of MBS~\cite{lutchyn.bakkers.18}. 
In the second case, MBS are expected due to the helical ordering of magnetic moments in the mono-atomic chains~\cite{braunecker.simon.13,klinovaja.stano.13,vazifeh.franz.13,braunecker.simon.15,kaladzhyan.simon.17,andolina.simon.17}.

MBS emerge in these systems when they are in a topologically non trivial phase. 
In a typical situation, the phase transition from topologically trivial to non trivial  is induced by the magnetic field~\cite{sato.takahashi.09,sato.fujimoto.09,sato.takahashi.10}.  
Increasing the applied magnetic field leads to a closing of the trivial superconducting gap and the reopening of a new, non trivial, gap~\cite{kobialka.ptok.19}. 
This is true for a system with a relatively small density of particles, i.e.~when the Fermi level is near the bottom of the band. 
If the splitting of the bands by the external magnetic field is larger than the superconducting gap, pairing occurs in the one--band channel~\cite{das.ronen.12,klinovaja.loss.12,fleckenstein.dominguez.18,kobialka.ptok.19}. 
This ``one'' type quasiparticle paring arises as an effect of the spin mixing by the SOC, corresponding to $p$-wave inter--site pairing in real space~\cite{gorkov.rashba.01,seo.han.12,ptok.rodriguez.18}. 

However, state of the art experiments also allow one to create inhomogeneous periodic magnetic fields.  For example, carbon nanotubes coupled to an antiferromagnetic substrate~\cite{desjardins.contamin.19} lead to a {\it synthetic magnetic field}~\cite{yazdani.19}.
Similar solutions were proposed theoretically in the form of nanomagnets~\cite{klinovaja.stano.12,kjaergaard.wolms.12,klinovaja.loss.13}, which have also been executed experimentally with an arrangement of alternating magnetization~\cite{maurer.gamble.18,sapkota.eley.19,kornich.vavilov.19}. 
Just like in the case of the magnetic moments with helical order~\cite{braunecker.simon.13,klinovaja.stano.13,vazifeh.franz.13,braunecker.simon.15}, this magnetic field can be the source of an effective spin--orbit coupling. 
Another possibility consists of magnetic nanopillars producing magnetic textures, which can be tuned by passing currents~\cite{zhou.mohanta.19}. 
Similar types of architecture based on magnetic tunnel junctions can be used to perform braiding operations~\cite{matosabiague.shabani.17}. 
Last, but not least, coupling a nanowire to a magnetic Co/Pt-multilayer~\cite{mohanta.zhou.19} can achieve a similar goal.

\begin{figure*}[!t]
\includegraphics[width=\textwidth]{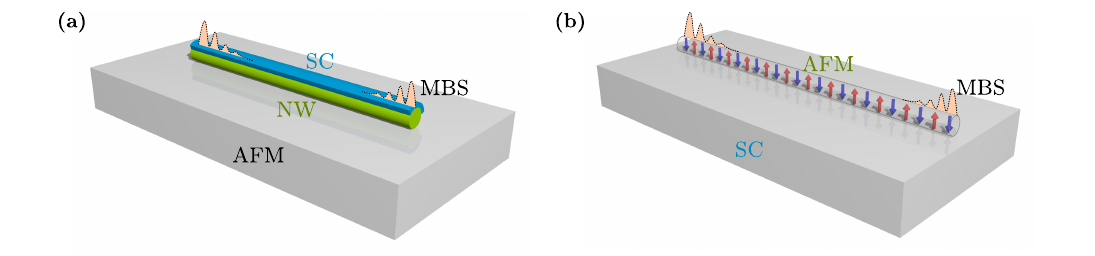}
\caption{A Schematic representation of the described systems -- in the case (a), a semiconducting nanowire (NW) is deposited on the surface of an antiferromagnetic (AFM) base, partially covered by a superconductor (SC); similarly, in  case (b) an AFM chain is deposited on a SC surface. 
In both cases, at the ends of the nanowire, Majorana bound states (MBS) can be induced due to the interplay between intrinsic spin orbit coupling, superconductivity, external Zeeman field (along the nanowire), and antiferromagnetism (induced by the proximity effect).}
\label{fig.schemat}
\end{figure*}

New perspectives for a system with antiferromagnetic (AFM) order were brought about by recent progress in experimental techniques allowing for the preparation of atomic chains~\cite{kim.palaciomorales.18}.
In such a case a self-organized spin helix order~\cite{braunecker.simon.13,klinovaja.stano.13,vazifeh.franz.13} can be stabilized via the Ruderman--Kittel--Kasuya--Yosida (RKKY) mechanism~\cite{zhou.wiebe.10,hermenau.brinker.19,meznel.mokrousov.12}.
Moreover, ideal monoatomic chains can be crafted with atoms one-by-one~\cite{kamlapure.corrnils.18}, which allows for the existence of various types of magnetic order in the chain~\cite{meznel.mokrousov.12,steinbrecher.rausch.18,schneider.brinker.20}. 
For example, AFM order was observed experimentally in sufficiently short Fe chains~\cite{loth.baumann.12,yan.malavolti.17,schneider.brinker.20} [c.f.~Fig.~\ref{fig.schemat}(b)].
Additionally, the proximity effect can relay the AFM order to the nanowire, e.g.~by contact with a strong AFM system [cf. Fig.~\ref{fig.schemat}(a)].
Strong antiferromagnets such as YbCo$_{2}$Si$_{2}$~\cite{pedrero.brando.10}, VBr$_{3}$~\cite{kong.guo.19}, Mn$_{2}$C~\cite{hu.wu.16}, NiPS$_{3}$~\cite{kim.lim.19}, or most promisingly V$_{5}$S$_{8}$~\cite{nakanishi.yoshimura.00,hardy.yuan.16}, can be  good candidates for the substrate in the investigated system. 
In such a case, topological phase can emerge due to the tuning of the external magnetic field, without destroying the AFM order in the substrate.

In the case of a semiconducting--superconducting hybrid nanowire, the non trivial topological phase is expected when the Fermi level is located near the bottom of the band.
Otherwise, a too large magnetic field is required.
As a result, the MBS is strongly restricted to the case of a low density of particles in the system. 
Contrary to this, we discuss a scenario for MBS in the nearly--half--filled case. 
The presence of MBS without any \emph{additional} external magnetic field applied, but instead only due to the AFM order, which we demonstrate here, has previously received only scant attention, see for example Refs.~\cite{heimes.kotetes.14} and \cite{heimes.mendler.15}.

This paper is organized as follows. 
In Sec.~\ref{sec.model}, we describe our model and the techniques used to investigate it. 
In Sec.~\ref{sec.topo_diag}, we derive the topological phase diagram of the system in the presence of AFM order and external magnetic field. 
We also discuss the origin of the non trivial topological phase.
In Sec.~\ref{sec.elec_prop}, we discuss electronic properties of the system in both real and reciprocal spaces.
Next, in Sec.~\ref{sec.diff}, we discuss the proposal of an experimental examination of this phase diagram via the differential conductance.
Finally, in Sec.~\ref{sec.sum}, we summarize the results.


\section{Model and techniques}
\label{sec.model}

\subsection{Real space description}
\label{sec.model_real}

\begin{figure}[!t]
\includegraphics[width=\columnwidth]{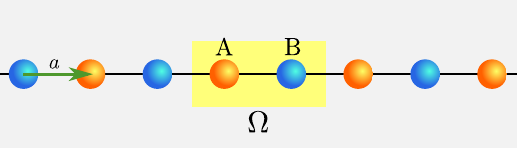}
\caption{The one dimensional AFM lattice discussed in this paper. The unit cell $\Omega$ contains two non-equivalent sites with opposite magnetic moments (orange and blue) belonging to sublattices A and B. The lattice spacing $a \equiv 1$ is taken as the distance between two nearest-neighbor sites.}
\label{fig.schemat2}
\end{figure}

In our calculations, we model the system shown schematically in Fig.~\ref{fig.schemat2}. We consider a one dimensional Rashba nanowire with superconducting and antiferromagnetic order both induced by proximity effects (cf.~Fig.~\ref{fig.schemat}), in the presence of an external magnetic field directed along the nanowire. 
The low energy physics of such a system can be described by the Hamiltonian $\mathcal{H} = \mathcal{H}_{0} + \mathcal{H}_\text{SC} + \mathcal{H}_\text{AFM}$.

The Rashba nanowire itself is described by
\begin{eqnarray}
\mathcal{H}_{0} &=& \sum_{ij,ss',\sigma} \left[ - t_{ij}^{ss'} - \left( \mu + \sigma h \right) \delta_{ij} \delta_{ss'} \right] c_{is\sigma}^{\dagger} c_{js'\sigma} \\
\nonumber &-& i \lambda \sum_{i,\sigma\sigma'} \left[ c_{iA\sigma}^{\dagger} \sigma^{y}_{\sigma\sigma'} c_{iB\sigma'} + c_{iB\sigma}^{\dagger} \sigma^{y}_{\sigma\sigma'} c_{i+1,A\sigma'} \right] + \textrm{H.c.} \,,
\end{eqnarray}
where $c_{is\sigma}^{\dagger}$ ($c_{is\sigma}$) describes the creation (annihilation) of an electron with spin $\sigma \in \{ \uparrow , \downarrow \}$ in sublattice $s \in \{ A , B \}$ of the {\it i}-th unit cell. We assume equal hopping between the nearest-neighbor sites (when $t_{ij}^{ss'} = t = 1$ in appropriate energy units) and zero otherwise. As usual, $\mu$ is the chemical potential, and $h$ is the external Zeeman magnetic field. In our calculations we neglect the orbital effect~\cite{kiczek.ptok.17}, assuming the magnetic field is parallel to the nanowire. The term in the second line describes the SOC with strength $\lambda$, where $\sigma_{y}$ is the second Pauli matrix. Superconductivity, which is induced in a nanowire due to the proximity effect, can be described by the BCS-like term:
\begin{eqnarray}
\mathcal{H}_\text{SC} &=& \Delta \sum_{is} \left( c_{is\uparrow}^{\dagger} c_{is\downarrow}^{\dagger} + c_{is\downarrow} c_{is\uparrow} \right) \,,
\end{eqnarray}
where $\Delta$ is the superconducting order parameter (SOP), proportional to the induced superconducting gap. The AFM order in the nanowire is described by
\begin{eqnarray}
\mathcal{H}_\text{AFM} &=& -m_{0} \sum_{i\sigma} \sigma \left( c_{iA\sigma}^{\dagger} c_{iA\sigma} - c_{iB\sigma}^{\dagger} c_{iB\sigma} \right) 
\end{eqnarray}
where $m_{0}$ denotes the amplitude of the AFM order.

\paragraph*{Finite size system.}---
Properties of the finite size system (with open boundary conditions), can be analyzed in real space. In this case, the Hamiltonian $\mathcal{H}$ can be diagonalized by the transformation $c_{is\sigma} = \sum_{n} \left( u_{isn\sigma} \gamma_{n} - \sigma v_{isn\sigma}^{\ast} \gamma_{n}^{\dagger} \right)$~\cite{degennes.89}, where $\gamma_{n}$ and $\gamma_{n}^{\dagger}$ are fermionic operators. Such a transformation leads to the real space Bogoliubov--de~Gennes (BdG) equations~\cite{balatsky.vekhter.06}, in the form $\mathcal{E}_{n} \Psi_{isn} = \mathbb{H}_{is,js'} \Psi_{js'n}$, where $\mathbb{H}_{is,js'}$ is the Hamiltonian in the matrix form, given in Appendix \ref{app:bdg}.

From solving the BdG equations, we can determine the site-dependent average number of particles:
\begin{eqnarray}
\label{eq.avarage_part} n_{is\sigma} &=& \langle c_{is\sigma}^{\dagger} c_{is\sigma} \rangle \\
\nonumber &=& \sum_{n} \left[ | u_{isn\sigma} |^{2} f \left( \mathcal{E}_{n} \right) + | v_{isn\sigma} |^{2} f \left( - \mathcal{E}_{n} \right) \right] \,,
\end{eqnarray}
where $f ( \omega ) = 1 / \left[ 1 + \exp \left( - \omega / k_{B} T \right) \right]$ is the Fermi--Dirac distribution.
From this, {\it effective} site-dependent magnetization is given as $m_{is} = n_{is\uparrow} - n_{is\downarrow}$.
In similar way, we can determine 
the local density of states (LDOS)~\cite{matsui.sato.03}:
\begin{eqnarray}
\label{eq.bdg_ldos} 
\rho_{is} (\omega) &=& - \frac{1}{\pi} \sum_{\sigma} \Imag \; G_{is\sigma} ( \omega + \im 0^{+} ) \\
\nonumber &=& \sum_{n\sigma} \left[ | u_{isn\sigma} |^{2} \delta(\omega + \mathcal{E}_{n}) + | v_{isn\sigma} |^{2} \delta(\omega - \mathcal{E}_{n}) \right] \,,
\end{eqnarray}
where $G_{is\sigma} = \langle c_{is\sigma} | ( \omega - H )^{-1} | c_{is\sigma}^{\dagger} \rangle$ and $\delta(\omega)$ is the Dirac delta function. 
The LDOS represents quantities experimentally measured by scanning tunnelling microscope (STM)~\cite{hofer.foster.03,wiesendanger.09,oka.brovko.14,stenger.stanescu.17}, and can give information about the emergence of the zero--energy states~\cite{nadjperge.drozdov.14}. 
In the numerical calculations, we replace the delta function by the Lorentzian $\delta(\omega) = \xi/[\pi (\omega^2+\xi^{2})]$, with a small broadening $\xi/t = 0.001$.

\subsection{Reciprocal space description}

From the explicit form the Fourier transform of operators:
\begin{eqnarray}
c_{is\sigma}^{\dagger}= \frac{1}{\sqrt{N}} \sum_{\bm k} c_{{\bm k}s\sigma}^{\dagger} \exp \left( - i {\bm k} \cdot {\bm R}_{is}\right) ,
\end{eqnarray}
where ${\bm R}_{is}$ denotes position of {\it i}-th sites in sublattice {\it s} (cf.~Fig.~\ref{fig.schemat2}), the Hamiltonian in momentum space can be found:
\begin{eqnarray}
\mathcal{H}_{0} &=& \sum_{{\bm k}\sigma} \mathcal{E}_{\bm k} \left( c_{{\bm k}A\sigma}^{\dagger} c_{{\bm k}B\sigma} + \text{H.c.} \right) \\
\nonumber &-& \sum_{{\bm k}s\sigma} \left( \mu - \sigma h \right) c_{{\bm k}s\sigma}^{\dagger} c_{{\bm k}s\sigma} \\
\nonumber &+& \sum_{{\bm k}\sigma\sigma'} i \mathcal{L}_{\bm k} \left(  c_{{\bm k}A\sigma}^{\dagger} \sigma_{\sigma\sigma'}^{y} c_{{\bm k}B\sigma'} + \text{H.c.} \right) \,,
\end{eqnarray}
\begin{eqnarray}
\mathcal{H}_\text{SC} &=& \Delta \sum_{{\bm k}s} \left( c_{{\bm k}s\uparrow}^{\dagger} c_{-{\bm k}s\downarrow}^{\dagger} + c_{-{\bm k}s\downarrow} c_{{\bm k}s\uparrow} \right) \,,
\end{eqnarray}
\begin{eqnarray}
\mathcal{H}_\text{AFM} &=& - m_{0} \sum_{{\bm k}s\sigma} \sigma \left( c_{{\bm k}A\sigma}^{\dagger} c_{{\bm k}A\sigma} -  c_{{\bm k}B\sigma}^{\dagger} c_{{\bm k}B\sigma} \right) \,,
\end{eqnarray}
where $c_{{\bm k}s\sigma}^{\dagger}$ ($c_{{\bm k}s\sigma}$) describes the creation (annihilation) operator of an electron with momentum ${\bm k}$ and spin~$\sigma$ in sublattice~$s$.
Additionally, $\mathcal{E}_{\bm k} = - 2 t \cos ( k )$ denotes the dispersion relation of non-interacting electrons in a 1D chain, while $\mathcal{L}_{\bm k} = - 2 i \lambda \sin ( k )$ is SOC in momentum space.

For the following we will use a more convenient representation for the Hamiltonian. We introduce Pauli matrices that act in the particle-hole subspace $\btau^{0,x,y,z}$, spin subspace $\bsig^{0,x,y,z}$, and sublattice subspace $\brho^{0,x,y,z}$. The ``$0$'' superscript labels the identity matrix for any given subspace. Then, following the Bogoliubov transform, the Hamiltonian in the Nambu basis,
\begin{eqnarray}
&& \psi_{\bm k}^{\dagger} = \\
\nonumber && \left( c_{{\bm k}A\uparrow}^{\dagger} \, c_{{\bm k}B\uparrow}^{\dagger} \, c_{{\bm k}A\downarrow}^{\dagger} \,  c_{{\bm k}B\downarrow}^{\dagger} \,  c_{-{\bm k}A\uparrow} \, c_{-{\bm k}B\uparrow} \, c_{-{\bm k}A\downarrow} \,  c_{-{\bm k}B\downarrow}  \right)\,,
\end{eqnarray}
takes the form $\mathcal{H} = \sum_{\bm k} \psi_{\bm k}^{\dagger} \mathcal{H} ( {\bm k} ) \psi_{\bm k}$, where
\begin{eqnarray}
\label{eq.ham_pauli} \mathcal{H} ( {\bm k} ) &=& \mathcal{E}_{\bm k} \btau^{z} \bsig^{0} \brho^{x} - \mu \btau^{z} \bsig^{0} \brho^{0} + i \mathcal{L}_{\bm k} \btau^{z} \bsig^{y} \brho^{x} \\
\nonumber &-& \Delta \btau^{y} \bsig^{y} \brho^{0} - h \btau^{z} \bsig^{z} \brho^{0} - m_{0} \btau^{z} \bsig^{z} \brho^{z} \,.
\end{eqnarray}
We will use this form of the Hamiltonian to calculate the bulk topological properties. 
In turn, due to the bulk--boundary correspondence~\cite{mong.shivamoggi.11,fukui.shiozaki.12}, this tells us when there will be MBS in the finite length nanowire. 
More details can be found in Sec.~\ref{sec.topo_diag}.

The band structure of the system can be found by diagonalizing the Hamiltonian~\eqref{eq.ham_pauli}. 
Each block $\mathcal{H}_{\bm k}$ has eight eigenvalues $\mathcal{E}_{\bm k}^{n}$ (for $n = 1,2,\ldots,8$) associated with eigenvectors
\begin{eqnarray}
\label{eq.bgd_pauli_eigenvec} & &\varphi_{\bm k} = \\
\nonumber && \left( u_{{\bm k}A\uparrow}^{n} \; u_{{\bm k}B\uparrow}^{n} \; u_{{\bm k}A\downarrow}^{n} \; u_{{\bm k}B\downarrow}^{n} \; v_{{\bm k}A\uparrow}^{n} \; v_{{\bm k}B\uparrow}^{n} \; v_{{\bm k}A\downarrow}^{n} \; v_{{\bm k}B\downarrow}^{n} \right)^{T} \,.
\end{eqnarray}

Due to the existence of the AFM order in the system, the unit cell $\Omega$ contains two non-equivalent sites.
Increasing the size of the unit cell twice leads to the folding of the the Brillouin zone (BZ) to ${\bm k} \in [ -\pi/2 , \pi/2 )$.
As a result the two time-reversal invariant momenta (TRIM)~\cite{moore.balents.07,dutreix.17} are ${\bm k} = 0$ and ${\bm k} = \pi/2$.
The impact of each parameter of the Hamiltonian on the band structure is described in detail in Appendix~\ref{app:band}).
As the AFM order introduces a band splitting at lower energies than in the standard scenario, we may expect that we can drive the chain into the non trivial phase at densities closer to the half-filling case.
As we shall see in the following, this is indeed the case.


\section{Topological phase diagram}
\label{sec.topo_diag}

In this section, we will discuss the topological phase diagrams obtained from analytical calculations of the invariants and numerical calculations. 
We will also consider them in the context of the localization of the Majorana zero modes at the ends of the system. 
Based on the symmetries of the system, we will discuss the origin of the topological phase and the impact of the AFM order.


\begin{figure}[!b]
\includegraphics[width=\columnwidth]{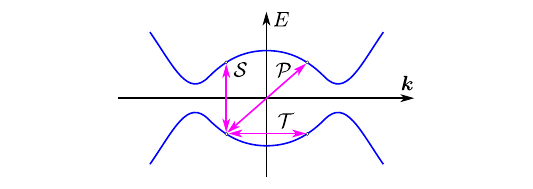}
\caption{
Schematic representation of the roles played by  the symmetries possessed by the considered Hamiltonian.
The particle--hole symmetry $\mathcal{P}$ results in the symmetry of the spectrum (solid blue line) with respect of the point `zero', while time--reversal symmetry $\mathcal{T}$ and chiral (sublattice) symmetry $\mathcal{S}$ correspond to reflection of the spectrum across the momentum and energy axes, respectively.
}
\label{fig.sym}
\end{figure}

\subsection{System symmetries}

The BdG Hamiltonian~(\ref{eq.ham_pauli}) can possess several symmetries important for its topological properties~\cite{sato.ando.17,chiu.teo.16}. Of interest are anti-unitary symmetries and we have:
\begin{enumerate}[(i)]
\item The particle--hole (PH) symmetry described by the anti-unitary operator $\mathcal{P} = \btau^{x} \bsig^{0} \brho^{0}\mathcal{K}$, such that $\mathcal{P} H_{\bm k} \mathcal{P}^{-1} = - H_{-{\bm k}}$ and $\mathcal{P}^{2} = 1$. $\mathcal{K}$ is the complex conjugation operator. It is worth mentioning, that all BdG Hamiltonians satisfy PH symmetry by construction~\cite{chiu.teo.16}.
\item The ``time-reversal'' (TR) symmetry described by the anti-unitary operator $\mathcal{T} = \lambda \mathcal{K}$, where $\lambda =\btau^{0} \bsig^{0} \brho^{0}$, and $\mathcal{T} H_{\bm k} \mathcal{T}^{-1} = H_{-{\bm k}}$ with $\mathcal{T}^{2} = 1$. Note that this is not the physical time-reversal operator for the electrons.
\item Finally we have the composite of these, the sublattice (SL) or ``chiral'' symmetry described by the unitary operator $\mathcal{S} =\mathcal{P}\mathcal{T}= \btau^{x} \bsig^{0} \brho^{0}$, with $\mathcal{S}^{-1} H_{\bm k} \mathcal{S} = - H_{\bm k}$.
\end{enumerate}
The impact of these symmetries on the Hamiltonian is schematically shown in the Fig.~\ref{fig.sym}.
With all of these symmetries present, which is the case for the Hamiltonian~(\ref{eq.ham_pauli}), we find ourselves in the BDI symmetry class in the Altland--Zirnbauer periodic-table of topological classes~\cite{altland.zirnbauer.97,ryu.schnyder.10,chiu.teo.16}.
From this, the $\mathbb{Z}$ invariant (i.e.~the winding number $w$) can be studied in order to discuss the topological phase diagram.
We can also construct a $\mathbb{Z}_{2}$ invariant (e.g.~from the Pfaffian) which measures the parity of $w$.
Both topological indices will be discussed below.

\subsection{Origin of the topological phase}

First we note that a Chiral Hamiltonian~(\ref{eq.ham_pauli}) can be rewritten in purely off-diagonal form~\cite{tewari.sau.12} using the rotation $\tilde{\mathcal{H}} ( {\bm k} ) = \mathcal{U}_{\pi}^{\dagger} \mathcal{H} ( {\bm k} ) \mathcal{U}_{\pi}$ where $\mathcal{U}_{\pi}=\e^{\im\frac{\pi}{4}\btau^y}\bsig^0\brho^0$. 
This results in
\begin{eqnarray}
\tilde{\mathcal{H}} ( {\bm k} ) &=& \mathcal{E}_{\bm k} \btau^{x} \bsig^{0} \brho^{x} - \mu \btau^{x} \bsig^{0} \brho^{0} + \im \mathcal{L}_{\bm k} \btau^{x} \bsig^{y} \brho^{x} \\
\nonumber &-&\Delta \btau^{y} \bsig^{y} \brho^{0} - h \btau^{x} \bsig^{z} \brho^{0} - m_{0} \btau^{x} \bsig^{z} \brho^{z}\,,
\end{eqnarray}
which has the form
\begin{eqnarray}
\label{eq.ham_offdiag} \tilde{\mathcal{H}} ( {\bm k} ) = \left(
\begin{array}{cc}
0 & \mathcal{A} ( {\bm k} ) \\ 
\mathcal{A}^{\dagger} ( - {\bm k} ) & 0
\end{array} 
\right) ,
\end{eqnarray}
where
\begin{eqnarray}
\label{eq.ak_form} \mathcal{A} ( {\bm k} ) &=& \mathcal{E}_{\bm k} \bsig^{0} \brho^{x} - \mu \bsig^{0} \brho^{0} + \im \mathcal{L}_{\bm k} \bsig^{y} \brho^{x} \\
\nonumber &+& \im\Delta \bsig^{y} \brho^{0} - h \bsig^{z} \brho^{0} - m_{0} \bsig^{z} \brho^{z} \,.
\end{eqnarray}
Now because
\begin{eqnarray}
\det \mathcal{H} ({\bm k}) = \det \tilde{\mathcal{H}} ({\bm k}) = \det \mathcal{A} ( {\bm k} ) \cdot \det \mathcal{A}^{\dagger} ( - {\bm k} ) ,
\end{eqnarray}
the sign of the gap is encoded by the function $Z_{\bm k} =  \det \mathcal{A} ( {\bm k} ) = \det \mathcal{A}^{\dagger} ( - {\bm k} )$, where from Eq.~(\ref{eq.ak_form}) we find
\begin{eqnarray}\label{eq.a_det}
Z_{\bm k}& =&
\left(h_{+}^2-\mu^2-\Delta^2\right)\left( h_{-}^2 - \mu^2 - \Delta^2 \right) \\
\nonumber &+& 8 t^2 \left( 2 t^2 \cos^2 (k) - h_{-} h_{+} - \mu^2 + \Delta^2 \right) \cos^2 (k) \\
\nonumber &+& 8 \lambda^2 \left( 2 \lambda^2 \sin^2 (k) + h_{-} h_{+} - \mu^2 + \Delta^2 \right) \sin^2 (k) \\
\nonumber &-& 16 t^2 \lambda^2 \sin^2 (k) \cos^2 (k) 
+32 \im t \Delta \lambda \mu \cos (k) \sin (k) ,
\end{eqnarray}
with $h_{\pm} = h \pm m_{0}$.

\begin{figure}[!b]
\includegraphics[width=\columnwidth]{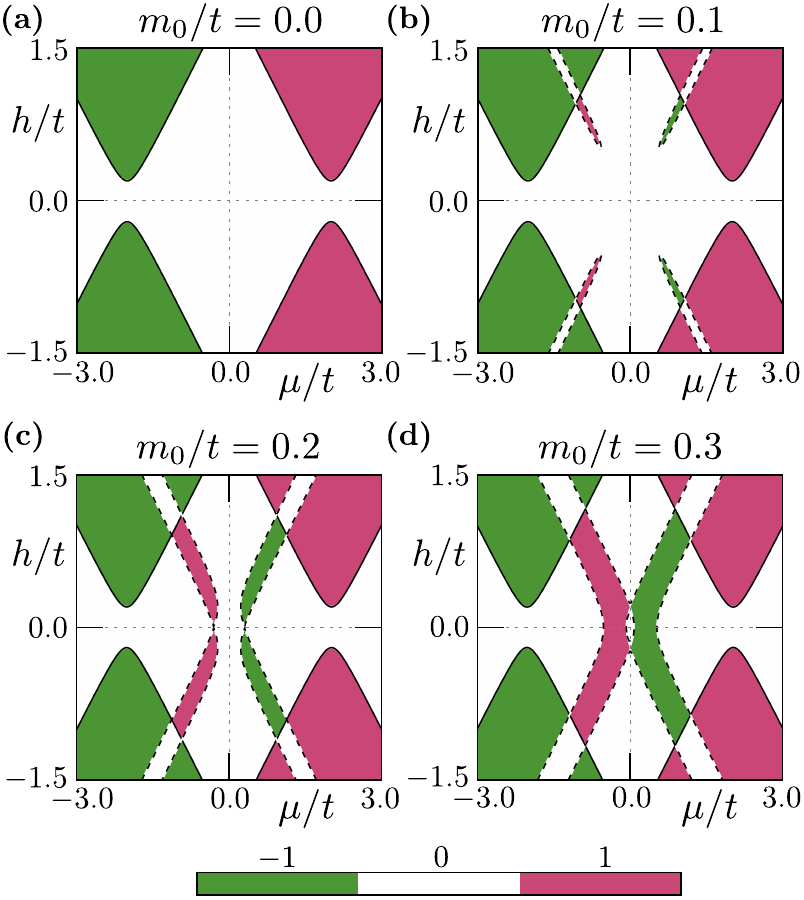}
\caption{
Topological phase diagrams obtained from the winding number $w$, given by Eq.~(\ref{eq.nu_inv}), for different amplitudes of the AFM order $m_{0}$ (as labeled).
Color denotes the trivial phase, $w=0$, (white) and the non trivial phases with $w=-1$ (green) and $w=1$ (red).
Solid black lines show gap closings at $k = 0$ and dashed lines show gap closings at $k = \pm \pi/2$.
Results are for $\Delta/t = 0.2$ and $\lambda/t = 0.15$.
}
\label{fig.diagram_nu}
\end{figure}

\begin{figure}[!b]
\includegraphics[width=\columnwidth]{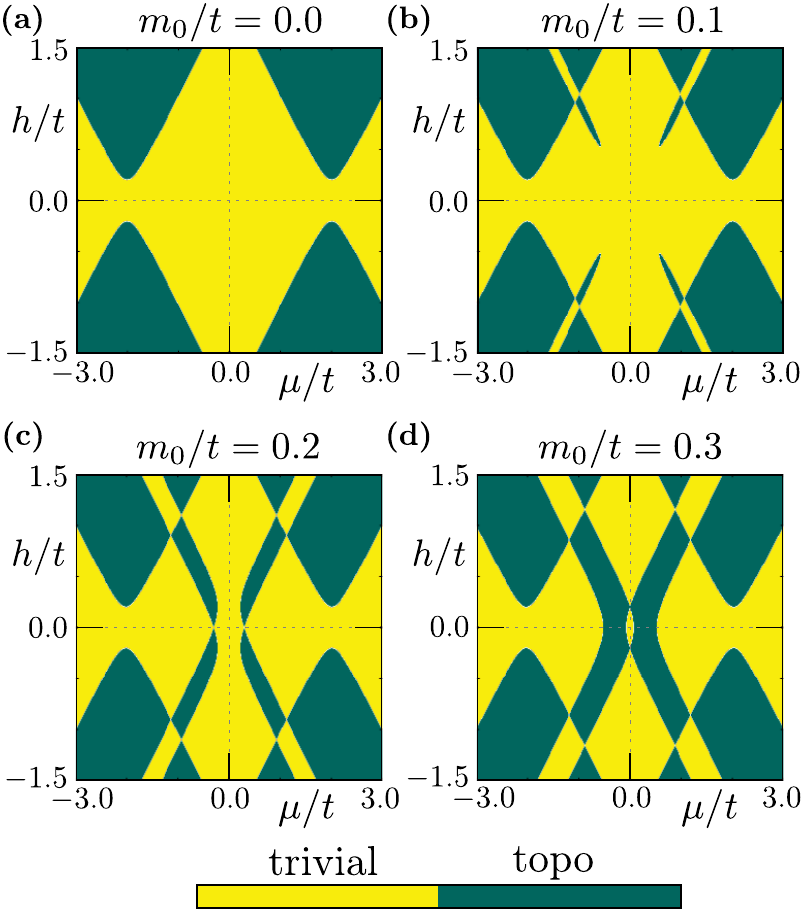}
\caption{
Topological phase diagrams, obtained from the topological index $
\mathcal{Q}$ calculated numerically using the $S$ matrix method (cf. Sec.~\ref{sec.smatrix}) for different amplitudes of the AFM order $m_{0}$ (as labeled).
Color denotes trivial (yellow) and non trivial (green) topological phase.
Results from $\Delta/t = 0.2$ and $\lambda/t = 0.15$ for lattice with $200$ sites without periodic boundary conditions.
}
\label{fig.diagram_smat}
\end{figure}

The non trivial topological phase can be found by calculating the topological invariant (see Appendix~\ref{app:topo} for more details).
In order to do that, we first define $z_{\bm k} = Z_{\bm k} / | Z_{\bm k} |$.
As $\mathcal{A} ( {\bm k})$ also has time reversal asymmetry $\mathcal{K}\mathcal{H}_{\bm k}\mathcal{K}=\mathcal{H}_{-{\bm k}}$, at the TRIM $\mathcal{A} (0,\pi/2)$ must be real, and hence so must $z_{0,\pi/2}$. Therefore for the topological index $w$ to change one of $z_{0,\pi/2}$ must pass through zero, corresponding to a gap closing. The relative signs of $z_{0,\pi/2}$ therefore encode some information about the topological index, its parity $(-1)^w$.
We can therefore construct a $\mathbb{Z}_{2}$ topological index~\cite{kitaev.01}:
\begin{eqnarray}
\label{eq.z2_index} \mathcal{Q} = ( -1 )^{w} =\sgn\left ( z_{{\bm k} = 0} \right) \cdot \sgn \left ( z_{{\bm k} = \pi/2 } \right)
\end{eqnarray}
which is equivalent to the index based on the Pfaffian~\cite{kitaev.01}
\begin{eqnarray}
\mathcal{Q} = \sgn \pf \left[ \mathcal{W} \left( 0 \right) \right] \cdot \sgn \pf \left[ \mathcal{W} \left( \pi/2 \right) \right] \,,
\end{eqnarray}
where $\mathcal{W}(k)=\mathcal{H}(k)\lambda$.
Moreover, from the Hamiltonian~(\ref{eq.ham_offdiag}) one finds:
\begin{eqnarray}
\nonumber \pf \left[ \mathcal{W} \left(0 \right) \right]&=&  
\left(h_{+}^2-\mu^2-\Delta^2\right)\left(h_{-}^2-\mu^2-\Delta^2\right)\\
&+& 8 t^2 \left( 2 t^2 - h_{-} h_{+} - \mu^2 + \Delta^2 \right) ,
\end{eqnarray}
and
\begin{eqnarray}
\nonumber \pf \left[ \mathcal{W} \left( \pi/2 \right) \right]&=&  
\left(h_{+}^2-\mu^2-\Delta^2\right)\left(h_{-}^2-\mu^2-\Delta^2\right)\\
&+& 8 \lambda^2 ( 2 \lambda^2 + h_{-} h_{+} - \mu^2 + \Delta^2 ) .
\end{eqnarray}
Topological phase diagrams obtained from Eq.~(\ref{eq.z2_index}) are in agreement with those ones obtained from the winding number (Fig.~\ref{fig.diagram_nu}), as well as from scattering matrix technique (Fig.~\ref{fig.diagram_smat}, cf.~Sec.~\ref{sec.smatrix}).

There exists a direct relation between the $\mathbb{Z}$ invariant $w$ and the Pfaffian $\mathbb{Z}_{2}$ invariant: $\mathcal{Q}=(-1)^w$~\cite{tewari.sau.12}.
As for this model $w\in\{-1,0,1\}$ then $\mathcal{Q}=1$ refers to a topologically trivial phase and $\mathcal{Q}=-1$ refers to a topologically non trivial phase.
It is then straightforward to find the exact relation between both invariants for our model:
\begin{eqnarray}
&& w = \\
&& \nonumber \frac{\sgn\left(\Delta\lambda\mu\right)}{2} \left\{
	\sgn \pf \left[ \mathcal{W} \left( \pi/2 \right) \right]-
	\sgn \pf \left[ \mathcal{W} \left(0 \right) \right]
	\right\} ,
\end{eqnarray}
which follows from Eq.~(\ref{eq.a_det}) and Eq.~(\ref{eq.a_inv}).
Topological phase diagrams obtained from the winding number calculations are shown in Fig.~\ref{fig.diagram_nu}.

Changes in $\mathcal{Q}$ are related to changes in the sign of $\pf \left[ \mathcal{W} \left( {\bm k} \right) \right]$ at TRIM.
In the absence of the AFM order ($h_{\pm} \rightarrow h$), only $\pf \left[ \mathcal{W} \left( 0 \right) \right]$ changes sign with changes in $h$.
This is shown as the typical form of the parabolic-like part on phase diagram [Fig.~\ref{fig.diagram_nu}(a)].
However, the existence of the AFM order alone can also force the emergence of an additional branch in the topologically non trivial phase.
This is possible due to the sign change of  $\pf \left[ \mathcal{W} \left( \pi/2 \right) \right]$ at the second TRIM $\pi/2$.
When the magnitude of the AFM order $m_{0}$ is significantly large, additional topological branches emerge from main branches (along $| \mu | \approx | h |$ line) [cf. Fig.~\ref{fig.diagram_nu}(a) and (b)].
If this occurs, for a range of parameters inside the main branches, the topological phase is destroyed as these phases have opposite chirality.
Further increasing of $m_{0}$ joins the AFM branches and leads to a destructive overlap and emergence of a trivial phase around $\mu = h = 0$ [Fig.~\ref{fig.diagram_nu}(d)].
When the AFM amplitude is relatively large, the non trivial phase can exist around $\mu \approx 0$, i.e.~in the nearly--half--filling limit $n \approx 1$ [cf.~Fig.~\ref{fig.diagram_nu}(c) and (d)].

Summarizing this part, the topological phase diagram is composed of two branches of the non trivial phase --- the main branch associated with TRIM at ${\bm k}=0$ and the additional branch connected with the second TRIM at ${\bm k}=\pm\pi/2$.
The main branch has properties which can be typically observed in the standard Rashba nanowire scenario, while the
non trivial phase originating in the additional branch can be compared to the non trivial phase induced by dimerization~\cite{kobialka.sedlmayr.19}.

\subsection{Scattering matrix method}
\label{sec.smatrix}

As an independent check of the preceding analytical calculations the behavior of the topological properties can be investigated by studying the scattering matrix $S$, which relates the incoming and outgoing wave amplitudes (further discussion on this point can be found in Sec.~\ref{sec.diff})~\cite{lesovik.sdovskyy.11,akhmerov.dahlhaus.11,beenakker.dahlhaus.11,rosdahl.vuik.18,fulga.hassler.11}.
In this method, the $\mathbb{Z}_2$ topological quantum number can be found from $\mathcal{Q} = \sgn \det R$, where $R$ denotes the reflection sub-matrix of $S$.
The scattering matrix can be calculated exactly from the real space Hamiltonian in the frame of the transfer--matrix scheme, described in detail in Ref.~\cite{fulga.hassler.11,choy.edge.11,zhang.nori.16}.
Using this method, we evaluated the topological phase diagram numerically.

Topological phase diagrams found with this method are shown in Fig.~\ref{fig.diagram_smat}.
The (non--)trivial topological phase covers the (green) yellow regions.
It can be seen that in the absence of the AFM order, the boundary of the non trivial phase in the $\mu$--$h$ space, is given by the known characteristic parabolas [Fig.~\ref{fig.diagram_smat}(a)].
The existence of the AFM order, modifies the boundaries of the non trivial phase around diagonal lines $|\mu| \approx |h|$ [Fig.~\ref{fig.diagram_smat}(b)], such a modification is a result of the presence of the sublattice in the system. 
These phase diagram were obtained numerically and are in complete agreement with the previous results obtained from analytical calculations (Fig.~\ref{fig.diagram_nu}).

\subsection{Topological phase without a Zeeman field}
\label{sec.mbz_nofield}

Analysis of these phase diagrams show important features of the described system:
first with the increase of the amplitude of $m_{0}$, we can see an emergence of additional \textit{branches} of non trivial phase.
Moreover, for some range of parameters the non trivial topological phase can emerge without any external magnetic field but instead, only due to the existence of the AFM order in the system.
This is manifested in the additional branch of the topological phase caused by the band inversion at the $ {\bf k} = \pi/2$ TRIM [cf.~Fig.~\ref{fig.diagram_nu}(d)].

Due to fact that the additional branch is connected with ${\bm k}=\pi/2$ TRIM, let us analyze the properties of $\pf \left[ \mathcal{W} \left( \pi/2 \right) \right]$ for $h=0$.
In this case $h_{\pm} = \pm m_{0}$, which gives
\begin{eqnarray}
\label{eq.h0_pf} \pf \left[ \mathcal{W} \left( \pi/2 \right) \right] \big|_{h=0} &=& \left[\Delta^2-m_0^2+\left(\mu+2\lambda\right)^2\right] \\
\nonumber &&\times\left[\Delta^2-m_0^2+\left(\mu-2\lambda\right)^2\right] .
\end{eqnarray}
One should note that in the limit $\lambda \rightarrow 0$, we have
\begin{eqnarray}
\pf \left[ \mathcal{W} \left( \pi/2 \right) \right] \rightarrow \left( m_{0}^{2} -  \mu^2 - \Delta^2  \right)^{2} \geq 0\,.
\end{eqnarray}
As we can see, SOC is still a mandatory ingredient of the non trivial topological phase.

\begin{figure}[!t]
\includegraphics[width=\columnwidth]{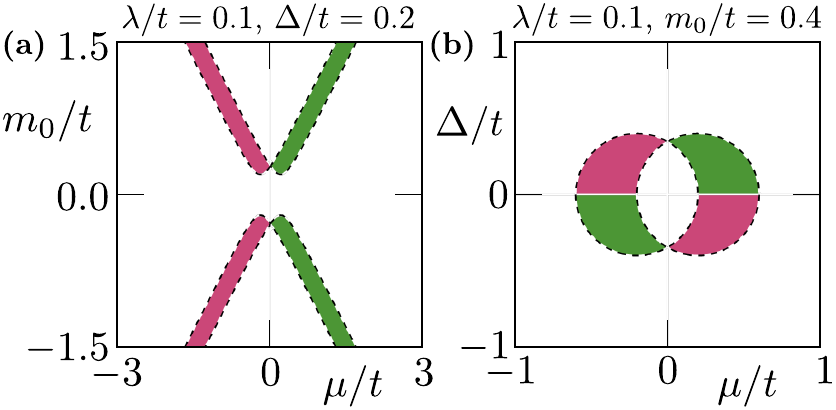}
\caption{
Topological phase diagram in the absence of the magnetic field.
Results are shown as a function of the amplitude of the AFM order and chemical potential (a); and of the SC order and and chemical potential (b).
The boundaries of the non trivial topological phases are given by the dashed lines.
The winding number $w$ color scheme is as in Fig.~\ref{fig.diagram_nu}.
In panel (b) the circles are centered on $\pm2\lambda$ with a radius $|m_0|$.
}
\label{fig.h0_phase}
\end{figure}

The impact of the AFM order amplitude and the SOC on the emergence of the non trivial topological phase is shown in  Fig.~\ref{fig.h0_phase}.
Interestingly, the boundaries of the non trivial topological phase are given exactly by two circles centred on $\pm2\lambda$ with a radius $|m_0|$ [Fig.~\ref{fig.h0_phase}(b)].
When the circles overlap each other, the overlapping region is in the trivial phase (no coloring).


\begin{figure}[!b]
\includegraphics[width=\columnwidth]{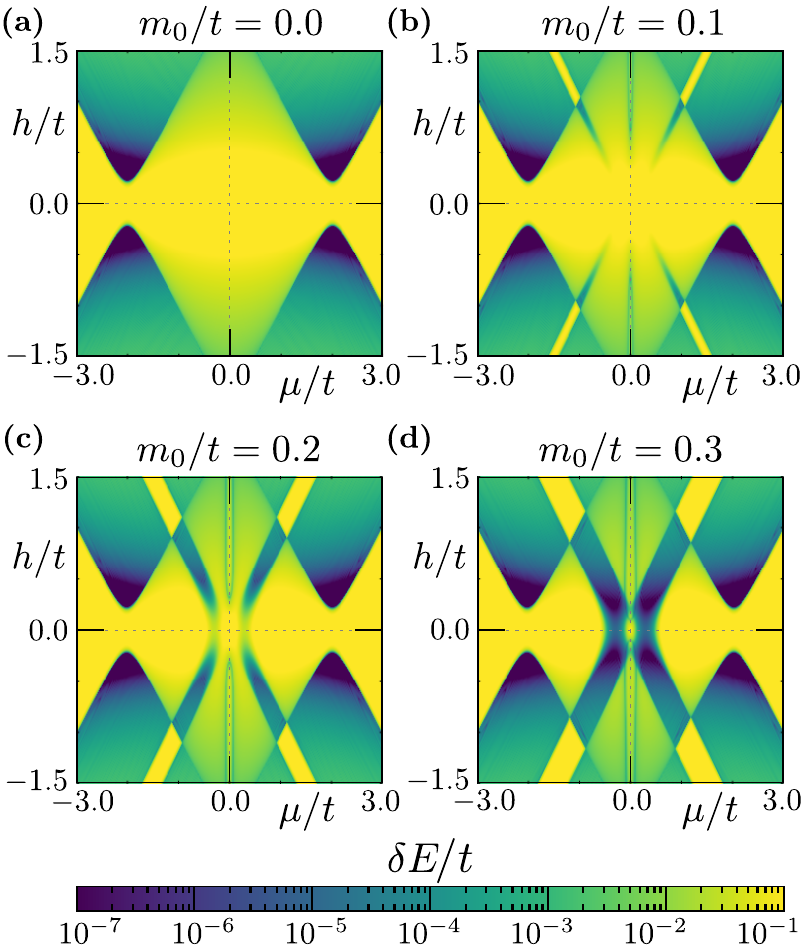}
\caption{
Values of the ``gap'' $\delta E$ defined as the difference between the energies of the two eigenstates which are nearest to the Fermi level for different amplitudes of the AFM order $m_{0}$ (as labeled).
}
\label{fig.diagram_warw}
\end{figure}

\begin{figure}[!b]
  \includegraphics[width=\columnwidth]{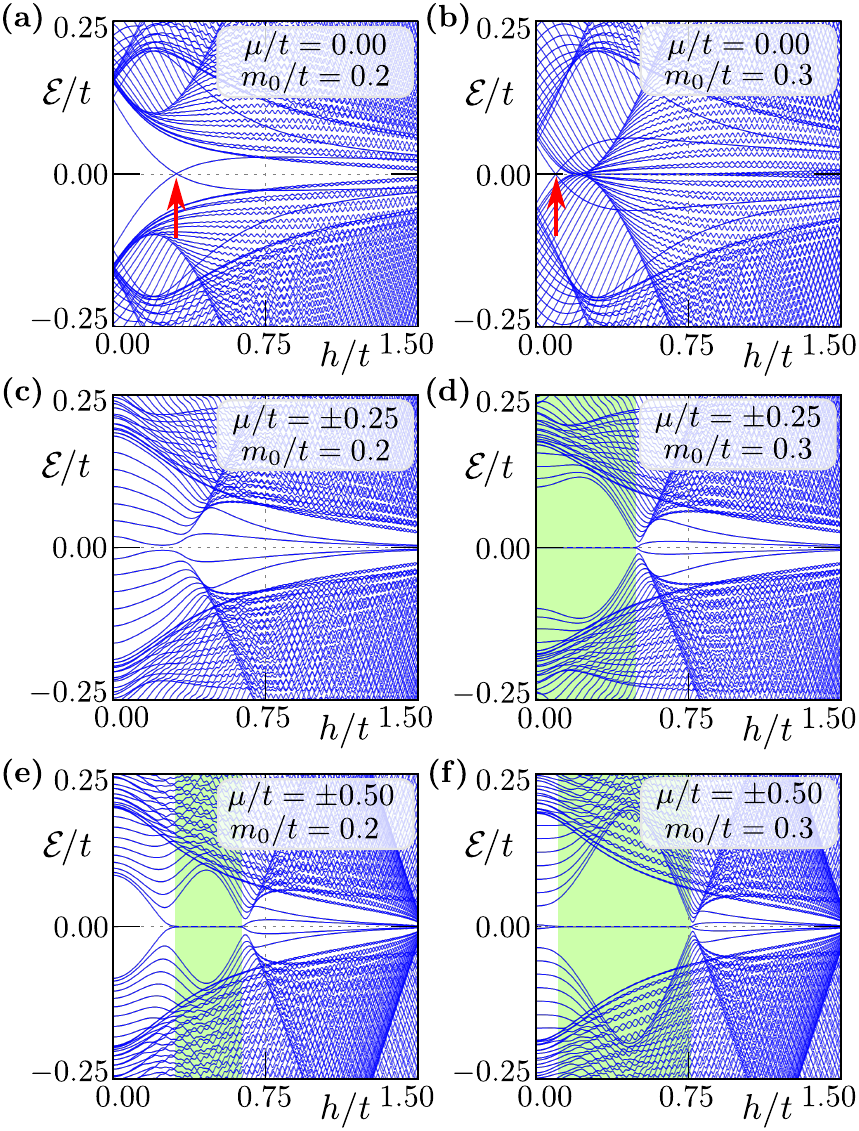}
\caption{
The spectrum of the system for various values of the chemical potential $\mu$ and the AFM amplitude $m_{0}$ (as labeled) as a function of the magnetic field $h$.
The range of $h$ marked by the green color corresponds to the non trivial topological phase.
}
\label{fig.eigen}
\end{figure}

\begin{figure}[!t]
  \includegraphics[width=\columnwidth]{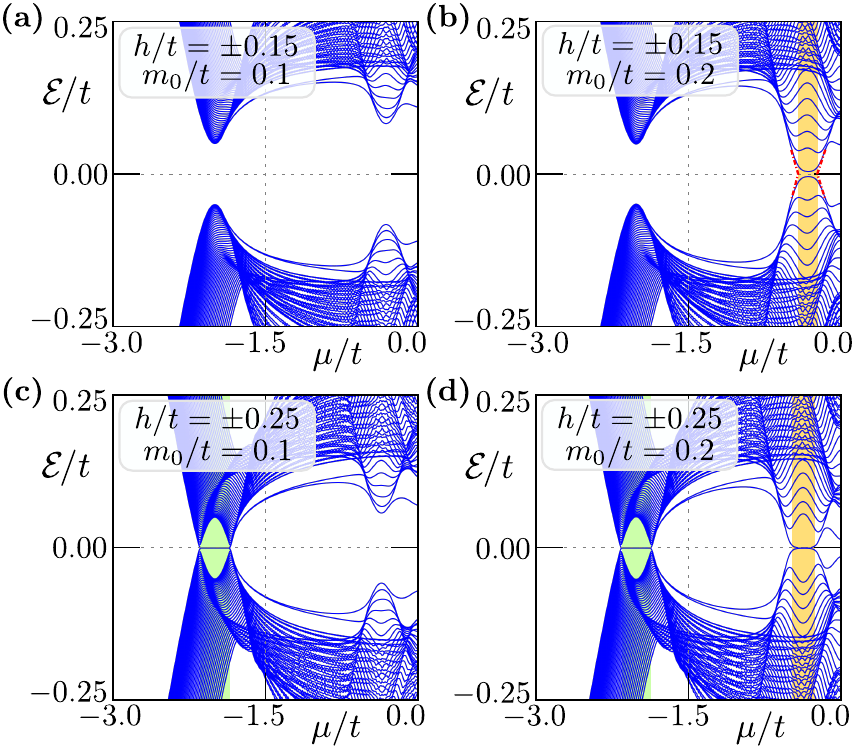}
\caption{
The spectrum of the system for various values of the magnetic field $h$ and the AFM amplitude $m_{0}$ (as labeled) as a function of the chemical potential $\mu$.
The range of $h$ is marked by the green and orange color, in correspondence with the non trivial topological phase in the main and additional branches respectively.
}
\label{fig.eigen_em}
\end{figure}

\section{Electronic properties}
\label{sec.elec_prop}

In this section, we will discuss the electronic properties of the system.
The numerical results presented in this section were obtained for a nanowire with with $N = 200$ sites and fixed values of $\Delta/t = 0.2$ and $\lambda/t = 0.15$.
Our tight binding parameters can be related to real quantities via $t = 1 / 2 m a^{2}$ and $\lambda \sim \alpha / a$, where $m$ is the electron's effective mass, $a$ is the lattice constant, and $\alpha$ is the physical spin--orbit coupling value.
However, any experimental realization of this system (c.f.~discussion in Sec.~\ref{intro}) can force particular system parameters.
For example~\cite{klinovaja.stano.13}, in the case of semiconducting wires $a \sim 0.6$~nm while $m = 0.027$~$m_\text{e}$, which gives $t \sim 10$~meV.
Induction of the superconducting gap by proximity effect is approximately given by $\Delta = 0.1$~meV. 
In this case, the Fermi level is located around the bottom of the band and can be tuned by doping or electrostatic gating.
Contrary to this, in the multiband monoatomic chains~\cite{kobialka.piekarz.19}, hoppings are in range of 0.5 eV.
Additionally $\Delta \simeq 1$~meV~\cite{klinovaja.stano.13}, while the Fermi level is located around half-filling.

\subsection{Topological gap and zero--energy states}
\label{sec.gap}

In the absence of symmetry breaking, a topological phase transition from a trivial to a non trivial phase is associated with closing of the trivial gap and reopening of a new topological gap.
In the case of the system without periodic boundary conditions, i.e.~with edges, the existence of MBS is equivalent to the existence of the nearly-zero-energy state after the phase transition to the non trivial topological phase.
However, a small value of the energy gap $\delta E$ (defined as a difference between energies nearest to the Fermi level in the spectrum of the system) is not a good indicator of the existence of MBS (Fig.~\ref{fig.diagram_warw}).
Still, a substantial decrease of $\delta E$ can indicate a clearly visible boundary between two topological phases, and has the advantage of being relatively straightforward to measure experimentally, in contrast to the invariants.
For instance, in the absence of the AFM order [Fig.~\ref{fig.diagram_warw}(a)], the phase boundary of the non trivial topological phase is visible in the form of characteristic parabolas.
An identical shape can be found in the corresponding topological phase diagram [cf. with Fig.~\ref{fig.diagram_nu}(a)].

The phase diagrams prove to be more complicated in the presence of the AFM order.
For some values of $\mu$, we can observe additional regions with extremely small values of $\delta E$, e.g.~vertical lines around $\mu/t = 0$ at Fig.~\ref{fig.diagram_warw}(c) and (d).
This behavior is associated with crossing of the Fermi level by the separate energy levels and can be noticed in the spectrum of the system [cf. red arrows at Fig.~\ref{fig.eigen}(a) and (b)].
Moreover, as these states exist in the trivial phase, they can not generate MBS at the end of the chain.

Energy spectra of the system are shown in Fig.~\ref{fig.eigen}.
For half--filling (i.e.~$\mu = 0$) some midgap states can cross the Fermi level $E = 0$ [shown by A red arrow in FigS.~\ref{fig.eigen}(a) and (b)].
However, a non trivial topological phase is not present and these states are not MBS.
In the non trivial topological phase, MBS are visible in the spectrum of the system in the form of two close to degenerate zero-energy states (the range of $h$ corresponding to the non trivial phase is marked by the green color in Fig.~\ref{fig.eigen}).
Similar to the nanowire without AFM order, eigenvalues show oscillations as a function of magnetic field $h$~\cite{dominguez.cayao.17,cayao.blackschaffer.18}.

Additional features of the system can be visible in the energy spectrum for constant field $h$ (Fig.~\ref{fig.eigen_em}).
The well known transition to the non trivial topological phase in the main branch is marked by the green areas.
The situation is more complicated in the case of the additional branch (marked by the orange areas in Fig.~\ref{fig.eigen_em}).
When $m_{0}$ is too small, the MBS do not fully emerge even if the non trivial topological phase due to finite size effects [cf. Fig.~\ref{fig.eigen_em}(b)].
Increasing $m_{0}$ increases the gap and the MBS can then form at the same system sizes [Fig.~\ref{fig.eigen_em}(d)].
In the case of the longer chains, the MBS exist even for smaller $m_{0}$ (cf.~Appendix~\ref{app:size}).
This shows the importance of the length scale $\zeta_{M}$ (denoting the exponential decay of the Majorana wavefunction in space~\cite{klinovaja.loss.12}), on the additional branch of the topological phase diagram.
The splitting of the MBS energy depends on the mutual relation between the chain length $L$ and $\zeta_{M}$, and in practice the emergence of a zero-energy MBS is possible when $L \gg \zeta_{M}$~\cite{zyuzin.rainis.13}.


\begin{figure}[!t]
\includegraphics[width=\columnwidth]{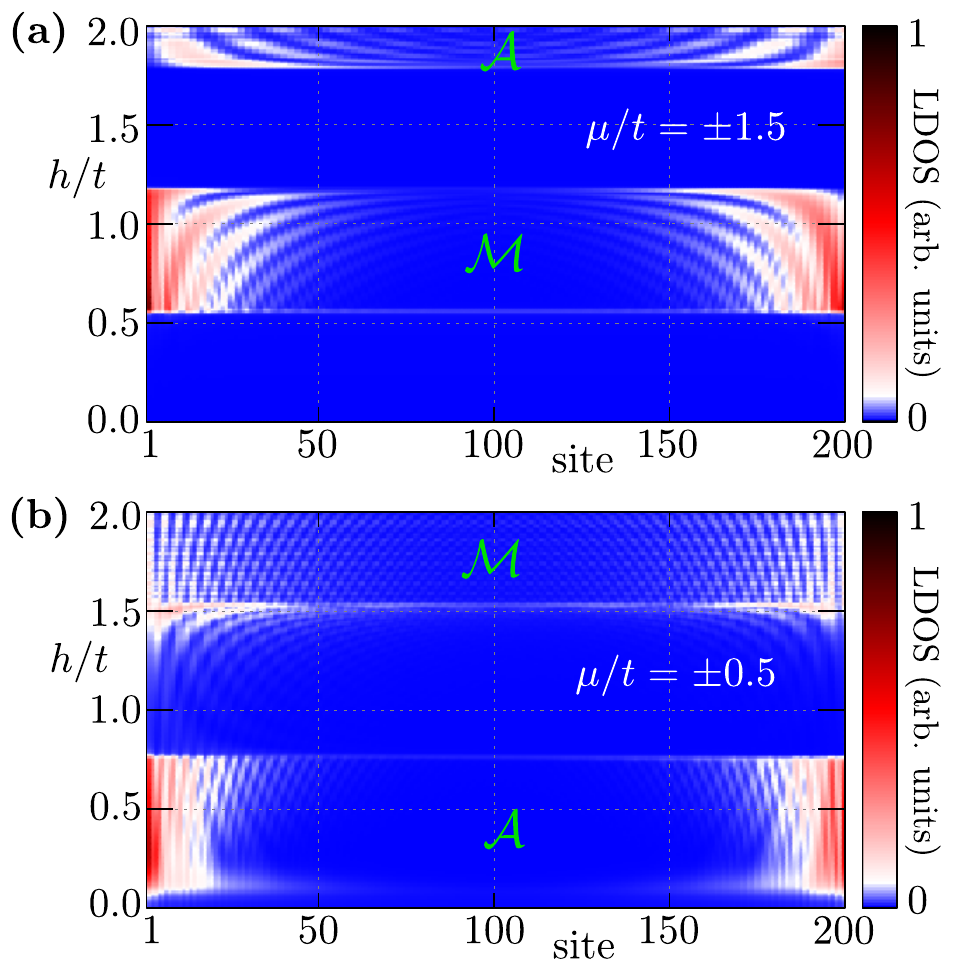}
\caption{
Zero-energy local density of states (LDOS) as a function of the magnetic field $h$.
Results are for AFM order with an amplitude $m_{0}/t = 0.3$.
Edge states localized within the main and additional topological branches are marked by $\mathcal{M}$ and $\mathcal{A}$, respectively.
}
\label{fig.ldos}
\end{figure}

\subsection{Localization of the Majorana modes}

Localization of the Majorana states can be studied via the zero--energy LDOS~(\ref{eq.bdg_ldos}).
Exemplary results for several values of the chemical potential $\mu$ as the magnetic field $h$ is increased are shown in Fig.~\ref{fig.ldos}.
The MBS emerging within the main and additional branches of the topological phase diagram are marked by $\mathcal{M}$ and $\mathcal{A}$, respectively.
In each topological phase, independently of the topological branch, the MBS are well localized around each end of the chain.
We can also observe the exponential decay of the MBS starting near the end of the chain and decaying to the middle.
Moreover, for sufficiently high $h$,  alternating oscillations of the Majorana bound states energies around the Fermi level are observed in the form of horizontal lines that represent the distribution of zero--energy states on the entire nanowire.
The same behavior have been discussed in the context of the energy gap in the previous section.
With increasing  magnetic field, we can observe a series of topological phase transition from trivial to non trivial and vice-versa as one would cross the branches of the topological phase diagram [cf.~Fig.~\ref{fig.diagram_nu}].

\subsection{Influence of the sublattices}

\begin{figure}[!t]
\includegraphics[width=\columnwidth]{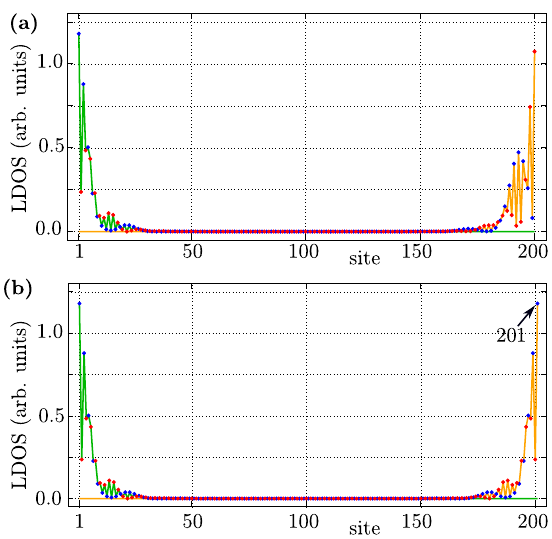}
\caption{
The zero-energy local density of states (LDOS) corresponding to the MBS. Blue and red dots represent the A and B sublattice sites, respectively. Results for $\mu/t = -0.3$, $h/t = 0.2$ and $m_{0}/t = 0.3$. A comparison of results for chains with (a) 200 and (b) 201 sites.
}
\label{fig.ldos_sub_size}
\end{figure}

From a diagonalization of the Hamiltonian in real space (cf.~Sec.~\ref{sec.model_real}) using the BdG formalism, in the non trivial topological phase we can find two zero-energy fermionic modes $\Psi^{\pm}$ at exponentially small energies $\pm\delta\epsilon$.
From this, using a simple rotation:
\begin{eqnarray}
\left(
\begin{array}{c}
\Psi_{is}^{L} \\ 
\Psi_{is}^{R}
\end{array} 
\right) = \frac{1}{\sqrt{2}}
\left(
\begin{array}{cc}
1 & 1 \\ 
-\im & \im
\end{array} 
\right)
\left(
\begin{array}{cc}
\Psi_{is}^{+} \\ 
\Psi_{is}^{-}
\end{array} 
\right) ,
\end{eqnarray}
we can find Majorana modes localized exactly at the left $\Psi^{L}$ or the right $\Psi^{R}$ side of the chain. Note that $\Psi^{L/R}$ are eigenstates of the particle-hole operator and therefore represent true Majorana modes.
In a similar way, using a site-dependent unitary transformation, we can find the representation of the Majorana wavefunction  $\Psi^{A/B}$ in each sublattice (A and B).
Exemplary results are shown in Fig.~\ref{fig.ldos_sub_size}, where left (right) modes are show by green (orange) solid lines, while the color of the dots (red and blue) represents the sublattices.

We will focus our analysis on the Majorana bound states manifesting in the additional branch of the topological phase diagram. 
In the absence of the external magnetic field the site-dependent distribution of the particles with opposite spins $n_{i\sigma}$ is given only by the AFM order. 
The total average number of particles per site $\langle n \rangle = \sum_{i\sigma} n_{i\sigma} / 2 N$ is always fixed by the chemical potential. 
The distribution of particles with spin $\uparrow$ and $\downarrow$ has a reflection symmetry with respect to the center of the system. 
Average number of particles in each site is approximately constant, however the AFM order introduced a distinguishability of the sublattices via magnetization---in other words magnetization in sublattice A is different to B. 
Additionally, the Majorana wavefunctions $\Psi^{L/R}$, as well as $\Psi^{A/B}$,  show reflection symmetry with respect to the center of the chain.
Here, it should be mentioned that the properties described above in the absence of the magnetic field do not depend on the parity of the number of sites.

\begin{figure}[!t]
\includegraphics[width=\columnwidth]{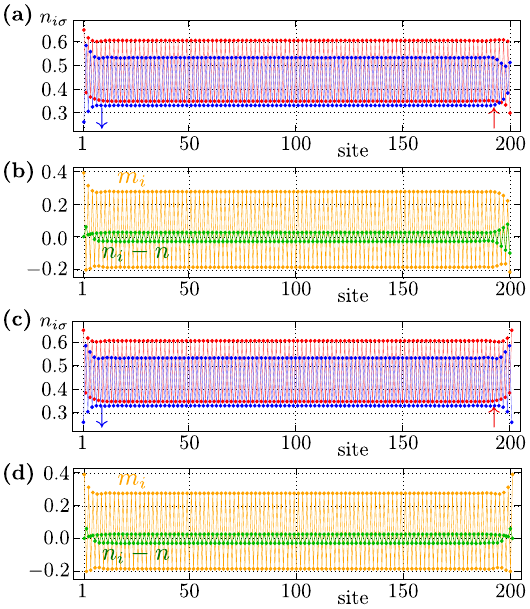}
\caption{
The distribution of particles $n_{i\sigma}$ with spin $\uparrow$ (red) and $\downarrow$ (blue) in the system, as well as magnetization $m_{i}$ (orange) and the difference between the average number of particles on a particular site and the average number per site in the system as a whole, $n_{i} - n$ (green).
A comparison of results for the chains with 200 [panels (a) and (b)] and 201 [panels (c) and (d)] sites is shown.
Same parameters as Fig.~\ref{fig.ldos_sub_size} for $\mu/t = -0.3$, $h/t = 0.2$ and $m_{0}/t = 0.3$.
}
\label{fig.ldos_sub_mn}
\end{figure}

The situation looks different in the presence of the magnetic field. 
In the case of the system with even number of sites [Fig.~\ref{fig.ldos_sub_size}(a)], this leads to a loss of the reflection symmetry.
This is a consequence of the modification of the $n_{i\sigma}$ distribution due to interplay between the AFM order and magnetic field.
In fact, the effective magnetic field $h_{\pm}$ at first and last site of the chain are not identical.
The reflection symmetry can be recovered by elongating the nanowire by one site, what yields an odd total number of  sites [Fig.~\ref{fig.ldos_sub_size}(b)]. 
As a result, first and last site belong to the same sublattice. 
Similar modification of the mirror symmetry by odd or even number of sites in the system is also observed in particle distributions. 
However, the bound states have only a very small influence on the particle distribution in the central region of the nanowire.

Properties similar to those described above are exhibited by the Majorana wavefunction.
When the number of sites is even [Fig.~\ref{fig.ldos_sub_size}(a)], the reflection symmetry is destroyed regardless of the basis ($\Psi^{L/R}$ or $\Psi^{A/B}$).
The reflection symmetry of the Majorana wavefunctions is restored when the system has odd sites [Fig.~\ref{fig.ldos_sub_size}(b)].
In this case, the $\Psi^{L}$ is a reflection of $\Psi^{R}$, while $\Psi^{A/B}$ has reflection symmetry with respect to the symmetry center.
Moreover, one of the $\Psi^{A/B}$ is greatly suppressed in one of the sublattices [in our case it is the sublattice marked by red points on Fig.~\ref{fig.ldos_sub_size}(b)].

\subsection{Spectral function analysis}

\begin{figure}[!t]
\includegraphics[width=\columnwidth]{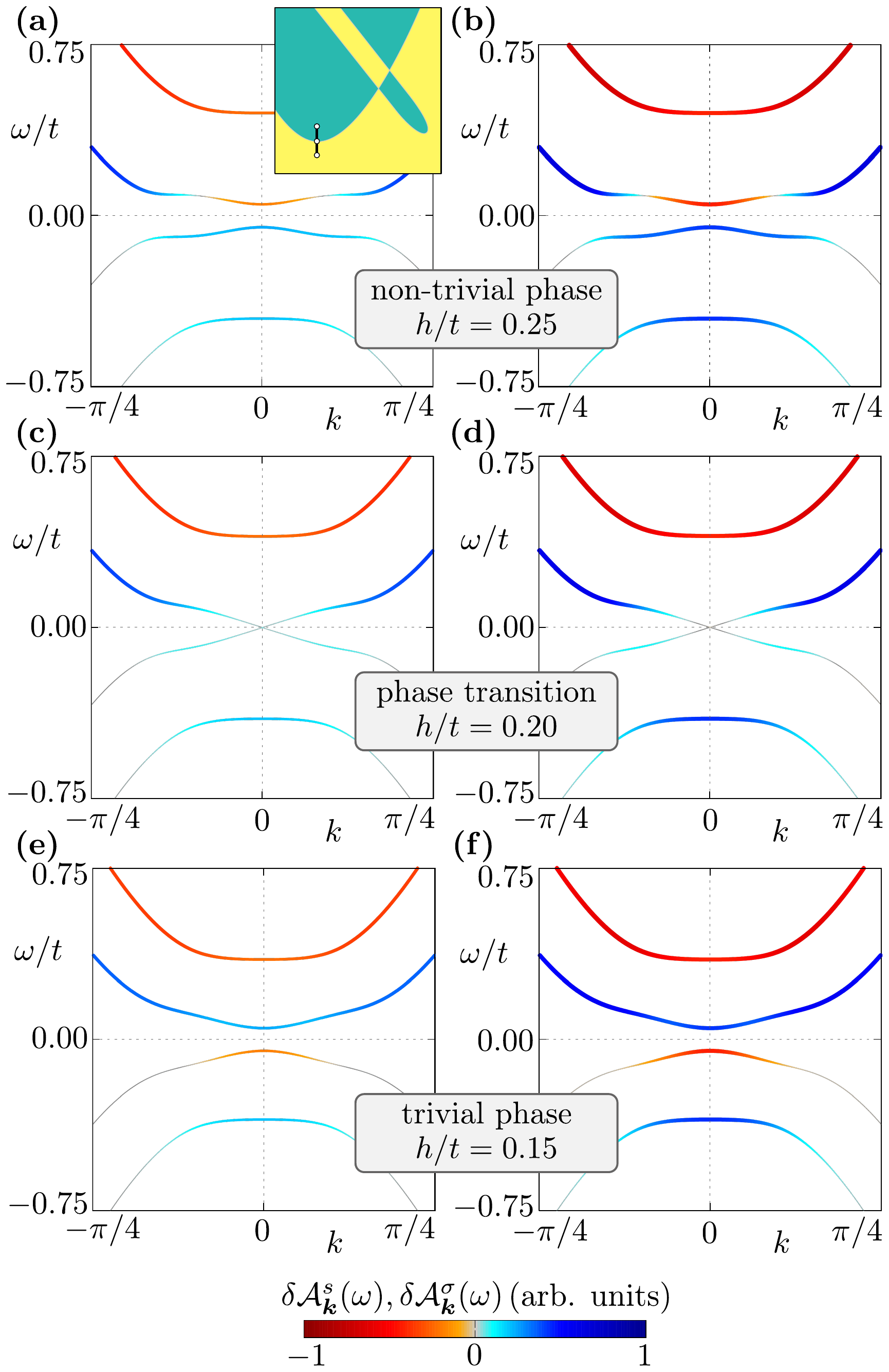}
\caption{
Distinctions in sublattice and spin dependent spectral functions, $\delta \mathcal{A}_{\bm k}^{s}$ (left panels) and $\delta \mathcal{A}_{\bm k}^{\sigma}$ (right panels).
Results for $m_{0}/t = 0.1$, $\mu/t = -2.0$ and $h/t=0.15$, $0.20$, and $0.25$ (panels from bottom to top) -- showing the topological phase transition from the trivial to the non trivial topological phase along the main branch (the black line is not in scale, shown in the inset).
}
\label{fig.bands_typical}
\end{figure}

\begin{figure}[!t]
\includegraphics[width=\columnwidth]{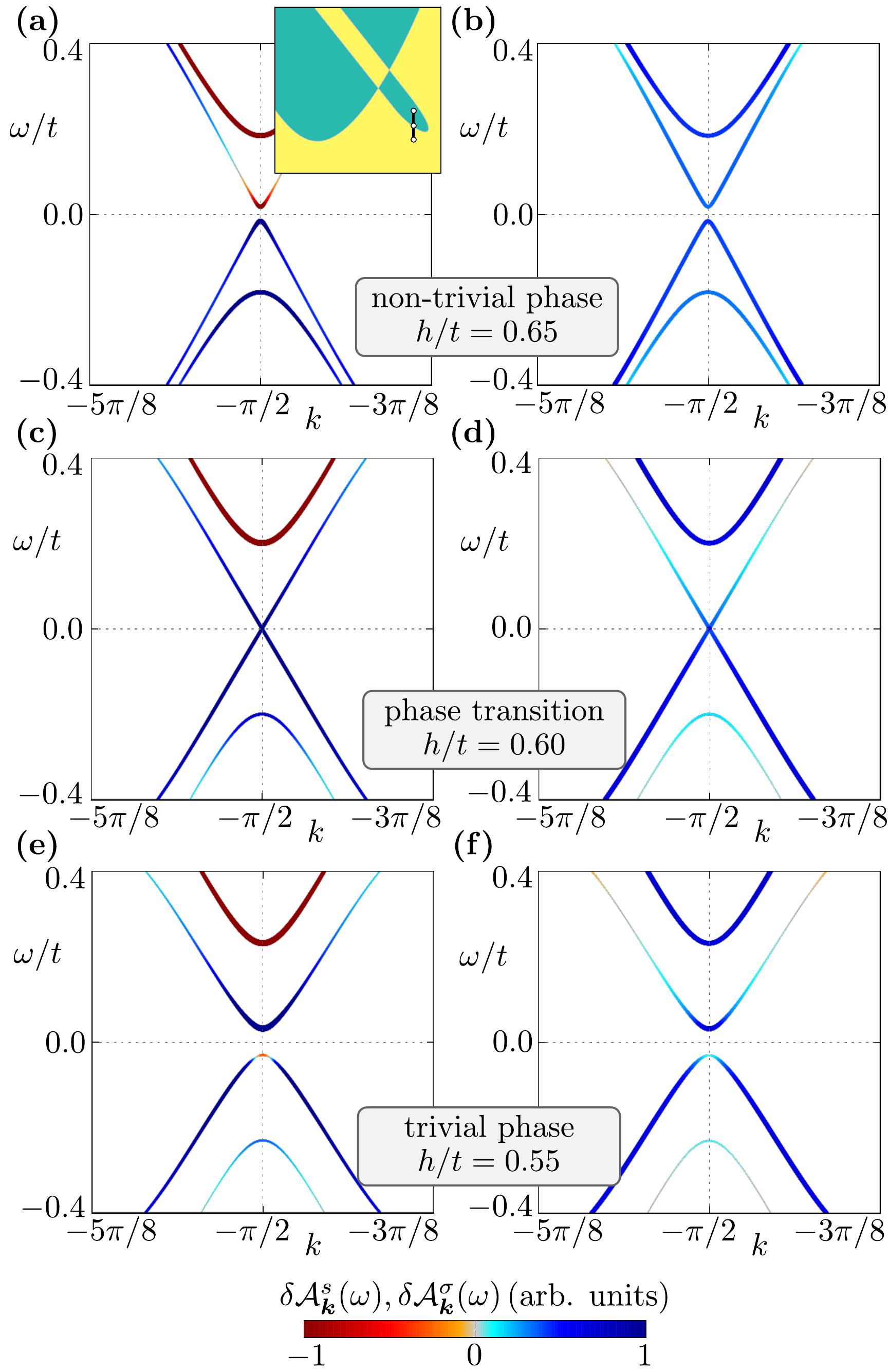}
\caption{
Distinctions in sublattice and spin dependent spectral function, $\delta \mathcal{A}_{\bm k}^{s}$ (left panels) and $\delta \mathcal{A}_{\bm k}^{\sigma}$ (right panels), respectively.
Results for $m_{0}/t = 0.1$, $\mu/t = -0.7$ and $h/t$ equal $0.55$, $0.60$, and $0.65$ (panels from bottom to top) -- topological phase transition from trivial to non trivial topological phase along additional branch (the green line is not in scale, shown at inset).
}
\label{fig.bands_untypical}
\end{figure}

The origin of the main and additional AFM branches in the topological phase diagram can be studied in the context of the spectral function:
\begin{eqnarray}
\mathcal{A}_{\bm k} ( \omega ) = - \frac{1}{\pi} \sum_{s\sigma} \Imag G_{{\bm k}s\sigma} ( \omega + \im 0^{+} ) \,,
\end{eqnarray}
where $G_{{\bm k}s\sigma} = \langle c_{{\bm k}s\sigma} | ( \omega - H )^{-1} | c_{{\bm k}s\sigma}^{\dagger} \rangle$.
In practice, the spectral function can be re-expressed in terms of the BdG coefficients:
\begin{eqnarray}
\mathcal{A}_{\bm k} ( \omega ) &=& \sum_{{\bm k}s\sigma} \mathcal{A}_{{\bm k}s\sigma} \\
\nonumber &=& \sum_{{\bm k}s\sigma} \left[ | u_{{\bm k}s\sigma}^{n} |^{2} \delta \left( \omega - \mathcal{E}_{{\bm k}n} \right) + | v_{{\bm k}s\sigma}^{n} |^{2} \delta \left( \omega + \mathcal{E}_{{\bm k}n} \right) \right] \,,
\end{eqnarray}
where $u_{{\bm k}s\sigma}^{n}$ and $v_{{\bm k}s\sigma}^{n}$ are components of the {\it n}-th eigenvector of the Hamiltonian~(\ref{eq.ham_pauli}).
Here we have introduced the sublattice- and spin-dependent spectral function $\mathcal{A}_{{\bm k}s\sigma}$.
The topological phase transition is associated with a band inversion during the transition.
To study this behavior in our system we can define
\begin{eqnarray}
\delta \mathcal{A}_{\bm k}^{s} &=& \mathcal{A}_{{\bm k}A\uparrow} + \mathcal{A}_{{\bm k}A\downarrow} - \mathcal{A}_{{\bm k}B\uparrow} - \mathcal{A}_{{\bm k}B\downarrow} , \\
\delta \mathcal{A}_{\bm k}^{\sigma} &=& \mathcal{A}_{{\bm k}A\uparrow} - \mathcal{A}_{{\bm k}A\downarrow} + \mathcal{A}_{{\bm k}B\uparrow} - \mathcal{A}_{{\bm k}B\downarrow} \,.
\end{eqnarray}
$\delta \mathcal{A}_{\bm k}^{s}$ and $\delta \mathcal{A}_{\bm k}^{\sigma}$ describe the imbalance in the sublattice and spin subspace at momentum ${\bm k}$ respectively.

Let us start with an analysis of the spectral function in the case when the topological phase arises in the main branch of the topological phase diagram (Fig~\ref{fig.bands_typical}).
As written previously, these topologically non trivial phases occur due to the band gap closing at the TRIM ${\bm k} = 0$.
Increasing the magnetic field, for fixed chemical potential, leads to a topological phase transition from the trivial to non trivial phase.
During this transition the band inversion is observed in both (sublattice and spin) subspaces.
Before the topological phase transition, i.e.~$h < h_{c}$, and in both subspaces, we observe the order of the bands ``polarization'': to be $(+,-,+,-)$, ordering from negative to positive energy [Fig~\ref{fig.bands_typical}(e) and (f)]. 
For the chosen parameters, topological phase transition occurs at the critical magnetic field $h_{c}/t \simeq 0.2$.
When $h = h_{c}$, the gap is closed and two bands touch each other at ${\bm k}=0$ [Fig~\ref{fig.bands_typical}(c) and (d)].
Further increase of $h$ leads to a changing of the ``polarization'' order to $(+,+,-,-)$ at ${\bm k}=0$ [Fig~\ref{fig.bands_typical}(a) and (b)]. At ${\bm k}=\pi/2$ the ordering remains $(+,-,+,-)$ and hence there is band inversion.
This inversion occurs in both sublattice and spin subspaces at the same time.
From this, we can conclude that the main branch of the topological phase emerges as an effect of the external magnetic field, independently of the AFM order.

Now, we turn to analyze the inversion of the bands in the case of the additional, AFM--related, branch of the topological phase diagram (Fig~\ref{fig.bands_untypical}).
In this case the existence of the topological phase is associated with the system properties at the TRIM ${\bm k} = \pm \pi / 2$.
As previously, increasing the magnetic field leads to the topological phase transition.
However, during this transition, in the spin sector we do not observe band inversion, i.e.~the spin polarization for each band is the same and positive (Fig~\ref{fig.bands_untypical} right panels) -- the spin imbalance in the system is unchanged due to the presence of a relatively strong magnetic field.
The situation looks different in the sublattice sector.
In the trivial phase [Fig~\ref{fig.bands_untypical}(e)], we observe band ordering as in the previous case, i.e.~$(+,-,+,-)$.
At $h = h_{c}$, we observe a closing of the gap at the TRIM ${\bm k}=\pi/2$ [Fig~\ref{fig.bands_untypical}(c)].
A further increase of $h$ leads to a band inversion in sublattice frame and the polarization order --- $(+,+,-,-)$ at ${\bm k}=\pi/2$ [Fig~\ref{fig.bands_untypical}(a)].
From this we can conclude, that the key role of AFM order is key in the emergence of the additional branch in the topological phase diagram.
Moreover, the introduction of the sublattice imbalance by the AFM order is the main source of the non trivial band topology.

Band inversion is a very typical signature of a topological phase transition in these systems~\cite{hasan.kane.10,bansil.lin.16,rider.palmer.19}, and was also reported as a signature of the topological phase transition in the case of the Rashba chain~\cite{setiawan.sengupta.15,szumniak.chevallier.17,kobialka.ptok.19}.
The spectral function can be measured in angle-resolved photoemission spectroscopy (ARPES) experiments~\cite{matsui.sato.03}.
The properties described above open a new way for the experimental examination of the construction of the additional topological branches in the AFM chain, and their comparison with the standard branch.

Summarizing, we would like to point out that the emergence of the topological branch in the phase diagram is a consequence of the band inversion located around half-filing ($\mu/t ~ 0$).
Here, we should remember that a similar behavior can also be observed in the systems exhibiting folding of the Brillouin zone due to an increase in the sites/atoms in the ``primitive'' unit cell~\cite{sedlmayr.aguiarhualde.16,kaladzhyan.bena.17,marra.cuoco.17,rex.gornyi.20,poyhonen.weststrom.14}.
Therefore, increasing the number of allowed subbands leads to more complicated forms of the topological phase diagram.
However, in contrary to those systems, in our case the magnetic order can lead to an emergence of MBS even in the absence of the external magnetic field.
This behavior can be crucial in the experimental realization of the MBS in the chains with chiral magnetic order~\cite{meznel.mokrousov.12,steinbrecher.rausch.18,kim.palaciomorales.18,schmitt.moras.19} or spin-block systems~\cite{herbrych.heverhagen.20}.


\section{Transport properties}
\label{sec.diff}

\begin{figure}[!b]
\includegraphics[width=\columnwidth]{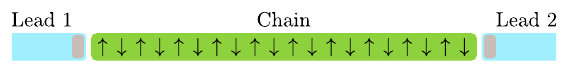}
\caption{
Schematic representation of system used in the differential conductance $G$ calculation --- AFM chain connected to two normal leads. 
Due to the Coulomb blockade between leads and chains, a barrier region exists in the system (gray area).
}
\label{fig.schemat3}
\end{figure}

Here, we show the results of numerical calculations of the differential conductance of the studied system using the scattering formalism~\cite{lesovik.sdovskyy.11,akhmerov.dahlhaus.11,beenakker.dahlhaus.11,fulga.hassler.11,rosdahl.vuik.18}. 
Our system, can be treated as a superconducting chain connected to normal leads (cf. Fig.~\ref{fig.schemat3}), i.e.~and N/S/N junction.
Then the scattering matrix relating all incident and outgoing modes in this system is:
\begin{eqnarray}
S = \left(
\begin{array}{cc}
S_{11} & S_{12} \\ 
S_{21} & S_{22}
\end{array} 
\right) , \quad S_{ij} = \left(
\begin{array}{cc}
S_{ij}^{ee} & S_{ij}^{eh} \\ 
S_{ij}^{he} & S_{ij}^{hh}
\end{array} 
\right) .
\end{eqnarray}
The $S_{ij}^{ab}$ is the block of scattering amplitudes of incident particles of type $b$ in lead $j$ to particles of type $a$ in lead $i$~\cite{rosdahl.vuik.18}.
The zero-temperature differential conductance matrix is
\begin{equation}
G_{ij} (E) \equiv \frac{\partial I_{i}}{\partial V_{j}} = G_{0} \left( T_{ij}^{ee} - T_{ij}^{he} - \delta_{ij} N_{i}^{e} \right) \,,
\end{equation}
where $I_{i}$ is the current entering terminal $i$ from the scattering region, while $V_{j}$ is the voltage applied to terminal $j$.
Here $G_{0} = e^{2} / \hbar$ is the conductance quantum without the spin degeneracy taken into account.
$N_{i}^{e}$ is the number of electron modes at energy $E$ in terminal $i$.
The energy transmission is given as
\begin{eqnarray}
T_{ij}^{ab} = \Tr \left( \left[ S_{ij}^{ab} \right]^{\dagger} S_{ij}^{ab} \right) \,.
\end{eqnarray}
We performed the calculation in the case of the N/S/N system shown in Fig.~\ref{fig.schemat3}, using the {\sc Kwant}~\cite{groth.wimmer.14} code to numerically obtain the scattering matrix.

\begin{figure}[!b]
  \includegraphics[width=\columnwidth]{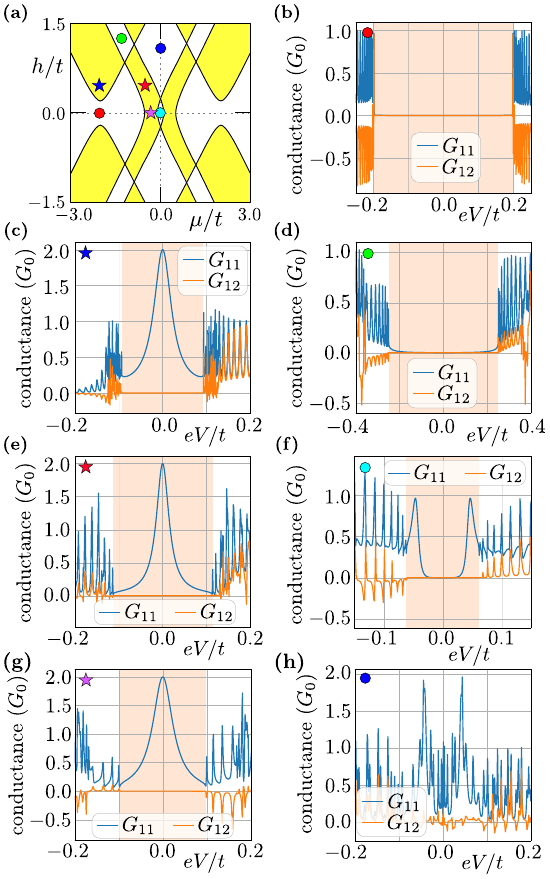}
\caption{
Local ($G_{11}$) and non-local ($G_{12}$) differential conductance  for different sets of the system parameters $\mu$ and $h$ (marked by colored points -- the stars and circles  correspond to the non trivial and trivial phases respectively).
Results for a finite size chain with $200$ sites and fixed $m_{0}/t = 0.3$, $\Delta/t = 0.2$ and $\lambda/t = 0.15$. 
}
\label{fig.cond}
\end{figure}

An experimental study of the MBS emergence in the system can be performed by local differential conductance $G_{ii}$ measurements (for $i = 1,2$).
In the tunnelling regime, the local conductance $G_{ii}$ in a normal lead probes the density of states in the proximitized region~\footnote{
We assume chemical potential $\mu/t = 0$ in leads and a barrier potential which is equal to $3 t$.
}.
From this, one can obtain information about the in-gap states close to the $i$-th normal lead.
In a typical situation, the local conductance $G_{ii}$ is quantized by $G_{0}$~\cite{wimmer.akhmerov.11} (if spin degeneracy is not present).
However, for ``true'' zero energy bound states, the local conductance $G_{ii}$ should be equal to $2G_{0}$ (per each MBS)~\cite{kjaergaard.nichele.16,zhang.liu.18,zhang.liu.19}.
A measurement of $G_{ii}$ in such a case can yield important information about the existence of the MBS and can be used in the experimental ``testing'' of the topological phase diagram~\cite{chene.yu.17}.
Contrary to this, non-local conductance $G_{12}$ (or $G_{21}$) can give information about the non trivial topological gap~\cite{rosdahl.vuik.18,ikegaya.asano.19} and be helpful in distinguishing between non trivial in-gap states and the ``bulk'' states.
The induced gap matches the energies at which the non-local conductance becomes finite~\cite{rosdahl.vuik.18}.

First, we evaluate  the local $G_{11}$ and non-local $G_{12}$ conductance for several fixed values of chemical potential $\mu$ and magnetic field $h$ (Fig.~\ref{fig.cond}). 
We assume $m_{0}/t = 0.3$, which corresponds to a rich topological phase diagram, cf.~Fig.~\ref{fig.cond}(a). 
In the simplest case, in the absence of the magnetic field, for chemical potential near the bottom of band ($\mu/t = -2$), i.e.~Fig.~\ref{fig.cond}(b), $G_{11}$ takes maximal values around $G_{0}$, while $G_{12}$ correctly shows the value of the gap (marked by the shaded orange background). 
The transition to the topological phase by increasing the magnetic field leads to the emergence of MBS associated with the zero-bias peak of $G_{11} = 2 G_{0}$, cf.~Fig.~\ref{fig.cond}(b). 
At the same time, non-zero value of $G_{12}$ show induced topological gap. 
In the intermediate trivial region, Fig.~\ref{fig.cond}(c) for $\mu/t = -1.5$ and $h/t = 1.25$, the results looks similar to the first case. 
Results obtained within the additional branch of the topological phase diagram, i.e.~Fig.~\ref{fig.cond}(e), look similar to the main branch -- $G_{12}$ indicate the values of the small topological gap with clearly visible zero-bias MBS peak $G_{11} = 2 G_{0}$. 
These features are also conserved in the absence of the external magnetic field [Fig.~\ref{fig.cond}(g), for $\mu/t = -0.25$].
Finally, in the central trivial region of the phase diagram, for $\mu/t=0$ and $h/t=0$, i.e.~Fig.~\ref{fig.cond}(f), again a typical signature of the trivial phase can be seen. 
Additionally, due to the closeness to the boundary of the topological phase, we observe a signature of the extremely small gap in $G_{12}$.
Similar behavior can be observed for larger values of $h$ [Fig.~\ref{fig.cond}(h), for $h/t = 1$], where in practice the gap is negligible.

\begin{figure}[!t]
  \includegraphics[width=\columnwidth]{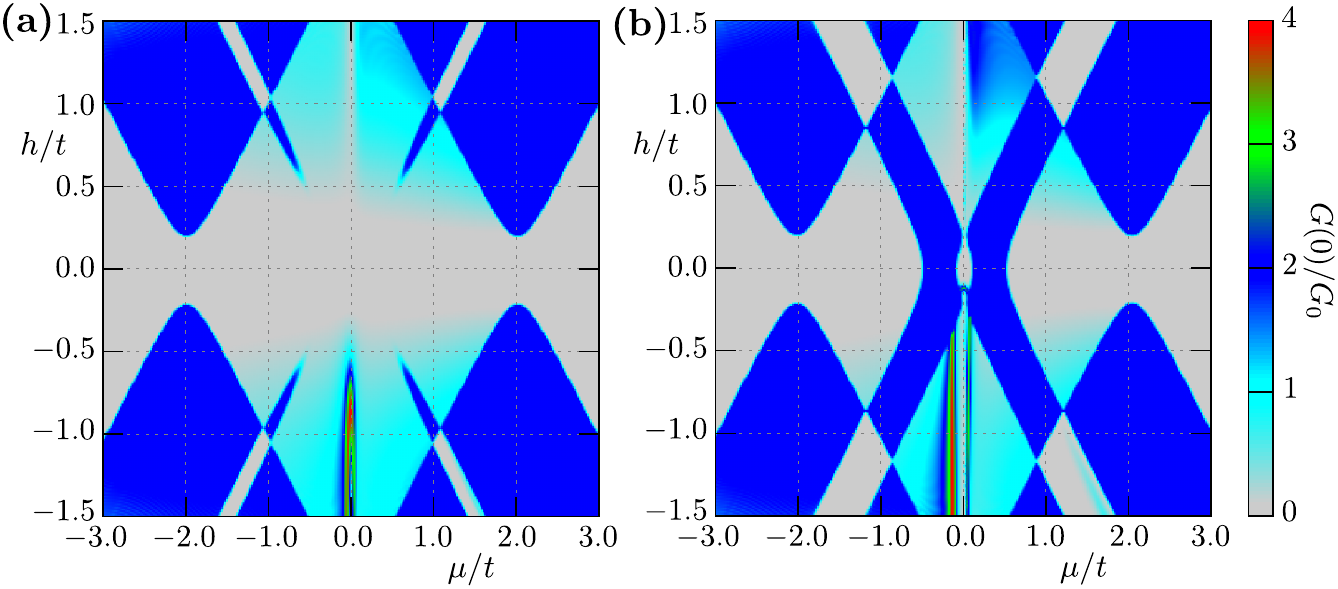}
\caption{
Value of the zero--bias local differential conductance $G_{11} (0)$ in the case of AFM nanowire with $m_{0}/t = 0.1$ (a) nad $m_{0}/t = 0.3$ (b).
Results for finite chain with $200$ sites, $
\Delta/t = 0.2$, and $\lambda/t = 0.15$.}
\label{fig.cond_ev0}
\end{figure}

Analogously to the experimental venue~\cite{chene.yu.17}, we can try to reproduce the shape of the topological phase diagram by studying the zero-bias local conductance $G_{11}$ (Fig.~\ref{fig.cond_ev0}). 
The conductance quanta $2G_{0}$ (the blue color) reproduce the main features of the topological phase diagram. 
As the calculations have been performed for a finite size system, in the case of a chain with 200 sites. 
For shorter chains  a vanishing of the MBS in some parts of the diagram can be observed. 
This effect is associated with the splitting of the in-gap energies~\cite{cao.zhang.19}, and is similar to the situation previously described in Sec.~\ref{sec.topo_diag}~A.


\subsection{Distinguishing trivial and topological zero energy states}

Although the existence of the $2G_{0}$ quantized conductance is often considered as a good indication of the presence of MBS, it should be noted that this can be mimicked by non-Majorana states~\cite{stanescu.tewari.19,moore.zeng.18}, and hence is not an unambiguous detection of a MBS. The non-local conductance $G_{ij}$ can give some information about the realization of the non trivial topological gap~\cite{rosdahl.vuik.18,ikegaya.asano.19}. 
A similar situation can also be found in hybrid systems, e.g.~in a nanowire with a quantum dot region, leading to the realization of non-topological zero-energy states~\cite{ptok.kobialka.17,reeg.dmytruk.18,moore.stanescu.18,moore.zeng.18}, which has also been reported experimentally~\cite{deng.vaitiekenas.16}.

Recently, there have been a host of methods introduced for distinguishing trivial zero-energy (Andreev or Yu--Shiba--Rushinov) bound states from the topological MBS. Those which may be directly applicable to the system we are considering include several theoretical predictions about non trivial spin signatures of MBS~\cite{sticlet.bena.12,haim.berg.15,bjornson.pershoguba.15,Guigou2016,szumniak.chevallier.17,kobialka.ptok.19} and spin selective Andreev reflection~\cite{he.ng.14,sun.zhang.16}. Such ideas were successfully applied within a spin polarized STM experiment as a diagnostic tool~\cite{jeon.xie.17}.
However due to the AFM background the spin polarization of the MBS is unlikely to show such clear results in this case.

Another way of observing a signature of MBS can be achieved, via coupling the topological nanowire to a quantum dot, by spin-resolved current shot-noise measurements~\cite{law.lee.09,haim.berg.15,liu.cheng.15,liu.levchenko.15,devillard.chevallier.17,smirnov.18} or finite-frequency current shot-noise~\cite{jonckheere.rech.20}.
An interesting alternative is possible due to the Majorana entropy study~\cite{smirnov.15}, which was successfully applied experimentally in the low temperature regime~\cite{hartman.olsen.18}.
The MBS may also be distinguished from other trivial bound states using supercurrents and critical currents measurements in superconductor-normal-superconductor junctions~\cite{cayao.blackschaffer.20,perrin.civelli.20,awoga.cayao.19}. These proposals all require significant modifications to the set-up under scrutiny here, and we will not consider them further in this work.


\section{Topological protection}
\label{sec.topo_prot}

\begin{figure*}[!t]
  \includegraphics[width=\linewidth]{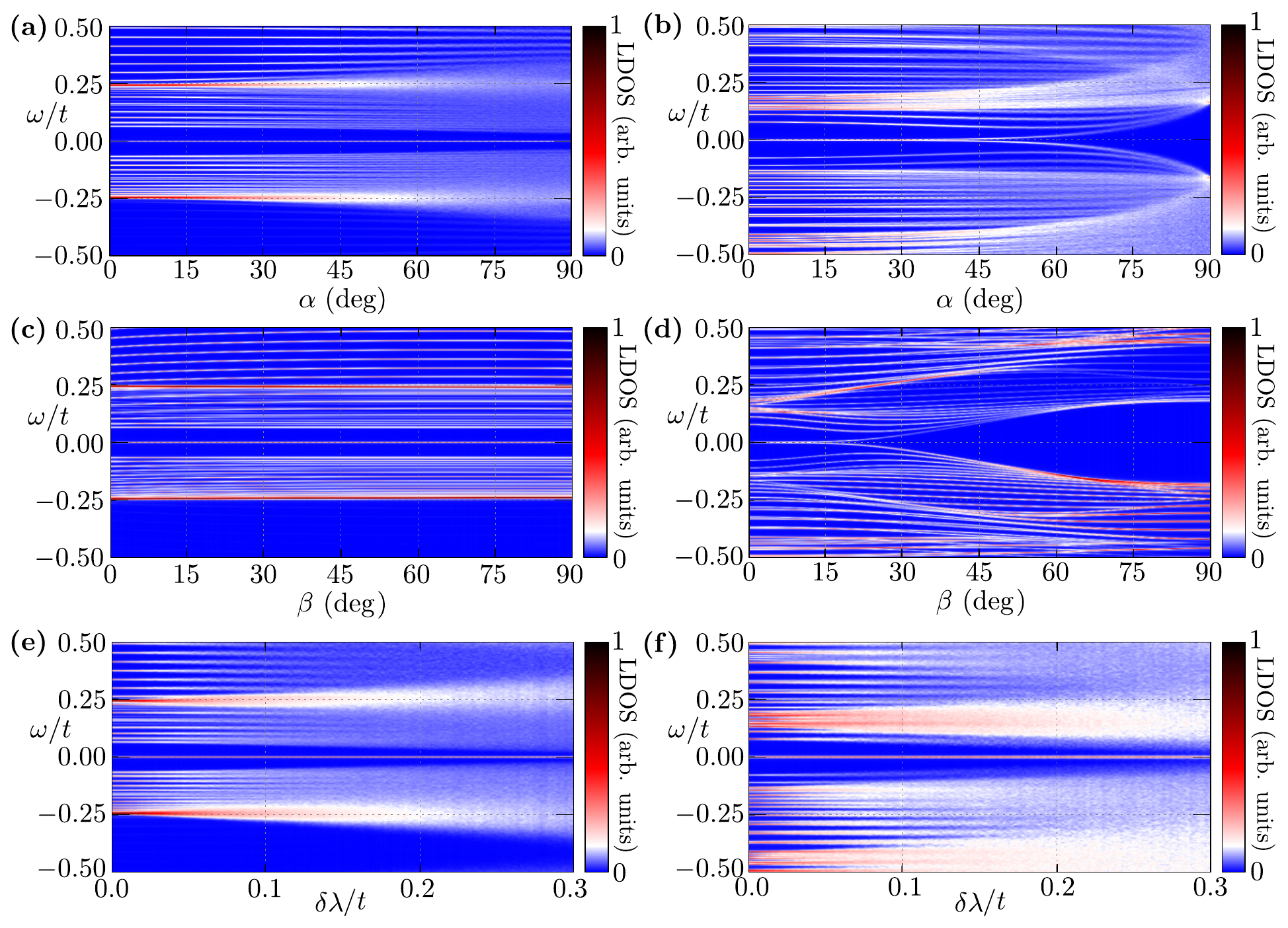}
\caption{
Comparison of the MBS stability due to different types of perturbation: a random tilt of the magnetic moments (a and b), a constant tilt of magnetic moments (c and d), and random perturbations of the SOC (e and f).
Degree value describes magnetic moments tilt from $\hat{e}_z$  to $\hat{e}_x$.
The color corresponds to the average LDOS under several configurations. 
Left and right panels show results within the main branch ($\mu/t = -2$, $h/t = 0.5$, and $m_{0}/t = 0.3$, blue star in~Fig.~\ref{fig.cond}) and the additional branch ($\mu/t = -0.2$, $h/t = 0.0$, and $m_{0}/t = 0.3$, violet star in~Fig.~\ref{fig.cond}), respectively.
}
\label{fig.disorder}
\end{figure*}

From a practical point of view, one of the most important properties of the MBS is their robustness due to the topological protection, which is manifested in the absence of an impact of any form of external ``disorder'' on the degeneration of MBS (provided the disorder neither closes the gap nor destroys the relevant symmetries).
We study this property in our system in the presence of several different types of perturbation: {\it (i)} a random, or {\it (ii)} homogeneous, tilt of the AFM magnetic moments; and {\it (iii)} random variations in the SOC coupling strength (i.e. off-diagonal disorder).
First, we modify the magnetic moment by a site-dependent perturbation $\delta m_{i}$ perpendicular to the initial AFM magnetic moments $m_{0}$.
We substitute:
\begin{eqnarray}
m_{0} \hat{e}_{z} \rightarrow \frac{ m_{0} \; \hat{e}_{z} + \delta m_{i} \; \hat{e}_{x} }{\sqrt{ 1 + \delta m_{i}^{2} / m_{0}^{2} }} ,
\end{eqnarray}
which conserves the norm of the magnetic moment on each site as equal to $m_{0}$.
In case {\it (i)}, $\delta m_i$ varies randomly for each site, whereas for {\it (ii)} the change in magnetization direction was homogeneous $\delta m_i\to\delta m$. 
For {\it (i)} we define the angle $\sin\alpha_i=\delta m_i/m_0$ and for {\it (ii)} we define the angle $\sin\beta=\delta m/m_0$.
Secondly, we assume for {\it (iii)} off-diagonal disorder as a perturbation of the SOC value:
\begin{eqnarray}
\lambda_{ij} \rightarrow \lambda + \delta \lambda_{ij} ,
\end{eqnarray}
where $\delta \lambda_{ij} = \delta \lambda_{ji}$ denotes the change in the SOC amplitude between neighbouring sites.

To study the influence of these perturbations on the robustness of the MBS, we calculated the DOS of the disordered system.
For cases  {\it (i)} and  {\it (iii)}, we average over $10^{2}$ different distribution of $\alpha_i$ and $\delta \lambda_{ij}$ respectively. The parameters vary such that $\alpha_i\in[-\alpha,\alpha]$ (analogically for angle $\beta$) and $\delta\lambda_{ij}\in[-\delta\lambda,\delta\lambda]$. We compare the effects of the perturbations for both a point in the main topologically non trivial phase, and the additional topologically non trivial phase (blue and violet stars in Fig.~\ref{fig.cond}, respectively).

As may be expected, MBS emerging within the main topological branch are stable to random variations in the AFM direction, case {\it (i)}, [Fig.~\ref{fig.disorder}(a)]. In contrast, if one is in the additional branch, which is related to the AFM order, one can see that the MBS are destroyed for large enough variations in the AFM field direction [Fig.~\ref{fig.disorder}(b)]. One can compare this to tilting of magnetic field in the normal nanowire set-up, which also destroys the topological phase~\cite{kiczek.ptok.17}. For case {\it (ii)}, again tilting the direction of the AFM order has no effect on the main topological phase [Fig.~\ref{fig.disorder}(c)]. For the additional branch of the phase diagram tilting the AFM order drives the system through a topological phase transition to a trivial phase [Fig.~\ref{fig.disorder}(d)]. This happens for a smaller value of $\beta$ than $\alpha$ [compare Figs.~\ref{fig.disorder}(b) and \ref{fig.disorder}(d)].

The situation is different in the case of the off-diagonal disorder [Fig.~\ref{fig.disorder}(e) and (f)].
For the main branch, Fig.~\ref{fig.disorder}(e), the value of the SOC is not important for the existence of the topological phase, and so the topological phase remains robust. One can see that eventually disorder will close the gap for sufficiently large values of $\delta\lambda$. The additional branch, Fig.~\ref{fig.disorder}(f), has a phase transition to the topologically trivial regime for $\lambda\approx0.21 t$, and $\lambda=0.15 t$ in the results of Fig.~\ref{fig.disorder}(f). We are therefore not far from the topological phase transition and may expect the disorder to fully close the gap. However, for the disorder values considered this has not yet occurred.
This situation is similar to the dimerized branch which occurs in a similar nanowire with SSH ordering, where the MBS should be destroyed when the amplitude of the perturbation is on the order of the hopping $t$~\cite{kobialka.sedlmayr.19}.

\section{Summary}
\label{sec.sum}

In this paper, we studied the possibility of the emergence of Majorana bound states in a nanowire with antiferromagnetic and superconducting order induced by proximity effects. We found that the topological phase diagram is composed of two branches of the non trivial topological phase. The main branch has the typical properties characteristic for a superconducting Rashba nanowire, while the second additional branch is associated with the existence of the antiferromagnetic order. Moreover, for some range of the parameters, the additional branch of the non trivial topological phase can ``survive'' even in the absence of the external magnetic field. In such a case, antiferromagnetic order is the source of the non trivial phase near the half--filling limit.

These results show an emergence of a new, antiferromagnetic topological phase that can be contrasted with the typical situation, when the Majorana bound states can emerge only if the density of the particles is sufficiently low (i.e.~when the Fermi level is located near the bottom of the band) and the system is under the effect of an external magnetic field. However, the phase transition to the non trivial topological phase can still induced by the external magnetic field or by changing of the chemical potential, i.e.~by doping. We show that the standard non trivial phase of such a nanowire has a different band inversion signature to that of the novel phase, which could be measured in ARPES experiments. We also explored experimental signatures of the MBS and topological gap in the local and non-local differential conductance.


\begin{figure*}[!t]
\includegraphics[width=\textwidth]{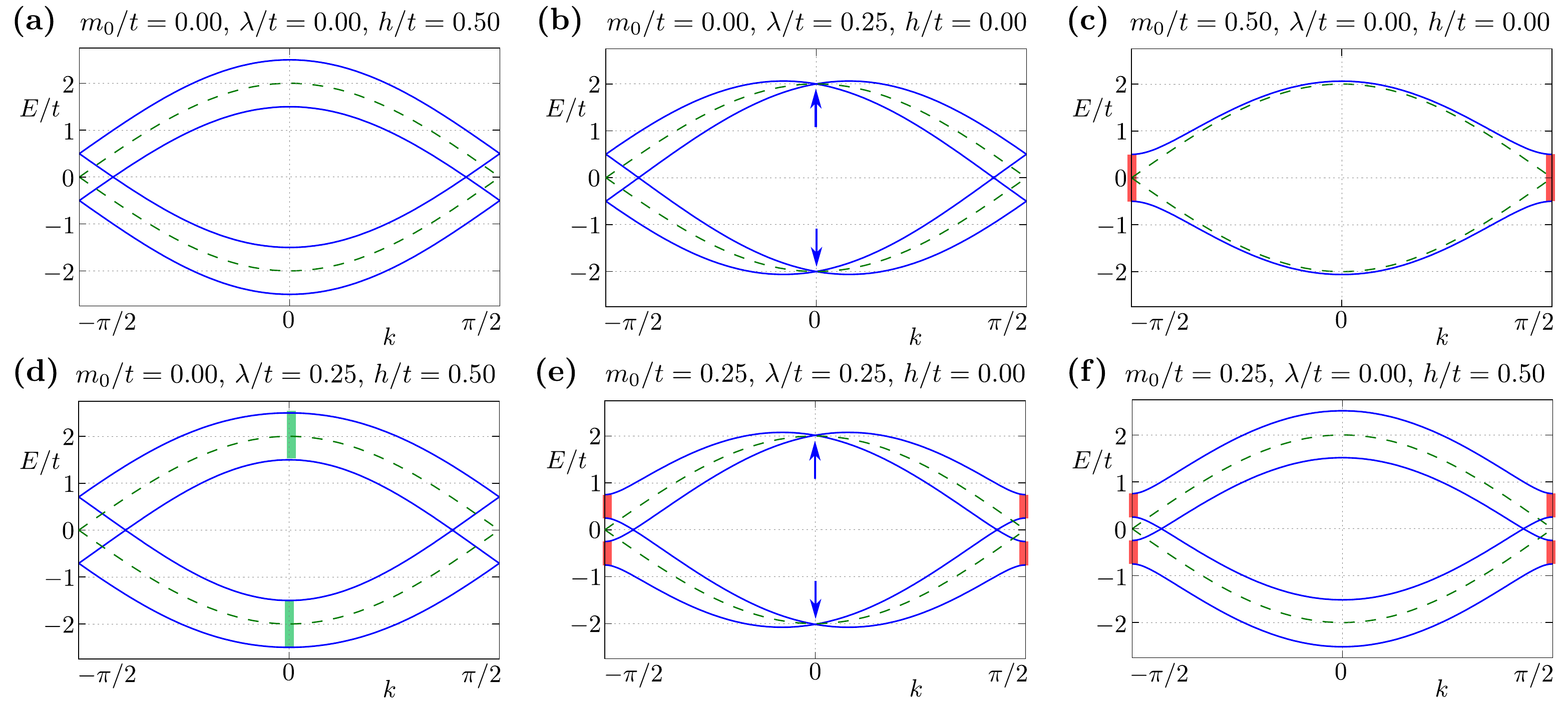}
\caption{
Impact of  the SOC and magnetic field on the spectrum of the chain without superconductivity. 
Results for several fixed parameters (as labelled) and chemical potential $\mu/t = 0$.
Dashed green line shows folding band for ``free'' chain with two sites in unit cell.
Green (red) markers at $k = 0$ ($k = \pm \pi/2$) indicate points with degeneracy lifted by the magnetic field (AFM order).
Results obtained for several fixed parameters (as labelled) and a chemical potential $\mu/t = 0$.
}
\label{fig.bands}
\end{figure*}

\begin{acknowledgments}
We kindly thank Pascal Simon for fruitful discussions.
This work was supported by the National Science Centre (NCN, Poland) under the grants 
UMO-2018/31/N/ST3/01746 (A.K.), 
UMO-2018/29/B/ST3/01892 (N.S.),
and 
UMO-2017/24/C/ST3/00276 (A.P.). 
Additionally, A.P. appreciates funding in the frame of scholarships of the Minister of Science and Higher Education (Poland) for outstanding young scientists (2019 edition, No.~818/STYP/14/2019).
\end{acknowledgments}

\appendix

\section{Band structure}
\label{app:band}

Here we discuss the impact of the model parameters on the band structure of the chain without superconductivity ($\Delta=0$).
The band structure is presented in Fig.~\ref{fig.bands}.
In the case of a ``free standing'' chain (i.e.~in the absence of magnetic field, AFM order, and SOC), the bands contain two spin degenerate branches due to the unit cell containing two, in this case identical, atoms -- upward and downward parabolic bands (dashed green line in every panel).
These two branches are a result of the folding of the $\mathcal{E}_{\bm k} = - 2 t \cos (k_{x}a)$ dispersion relation, intersecting at ${\bm k} = \pm \pi/2$.
The external magnetic field $h$ leads to a shifting of the bands in the energy domain due to the Zeeman effect [Fig.~\ref{fig.bands}(a)].
The Rashba type SOC leads to a shifting of bands in the momentum domain,, while preserving the band degeneracy at $k = 0$ (indicated by blue arrows) [Fig.~\ref{fig.bands}(b)].
Here, it should be mentioned that this effect is typical in the Rashba chain~\cite{bercioux.lucignano.15,manchon.koo.15}.
Introduction of AFM order into the system allows for band gap to emerge  at ${\bm k} = \pi/2$ [Fig.~\ref{fig.bands}(c)], marked by red background color.
This behavior has been also reported in the case of the Su--Schrieffer--Heeger (SSH) model, with two nonequivalent hopping between sites in unit cell~\cite{asboth.oroszlany.16}.
Such a band gap exists in the band structure independently of $\mu$ and other parameter [cf.~Fig.~\ref{fig.bands}(c), (d), and (f)].
Here, the spin degree of freedom remains a good quantum number, however the sublattice degree of freedom does not~\cite{baltz.manchon.18}.
This yields a situation where eigenstates are a spin-dependent mixture of the A and B sublattice states.
Moreover, the spatial profile displays a lattice dependent modulation of the density that is spin dependent and band dependent.
Breaking time reversal symmetry due to the AFM order still provides an analogue to Kramers' theorem due to the combined time reversal and translation symmetry --- hence there are two degenerate bands with opposite spins.
As a result, this degeneracy can be lifted by an external magnetic field (or by SOC).

Inclusion of such terms in pairs leads to a mixing of the aforementioned, separate, behaviors.
First, a magnetic field in the presence of SOC leads to a lifting of the band degeneracy at the $\Gamma$ point [indicated by the green markers in Fig.~\ref{fig.bands}(d), c.f.~with Fig.~\ref{fig.bands}(b)].
Second, AFM order and SOC shifts bands along the $k$ axis, while the band gap changes along the $E$ axis [Fig.~\ref{fig.bands}(e)].
At the same time, the degeneracy at ${\bm k} = 0$ (indicated by blue arrows) is preserved.
Still, a very strong magnetic field can lift this degeneracy (not shown).
Finally, the external magnetic field in the presence of the AFM order lifts the spin-degeneracy while simultaneously preserving the band gap at ${\bm k} = \pm \pi/2$ [Fig.~\ref{fig.bands}(f)].

\begin{figure}[!t]
\includegraphics[width=\columnwidth]{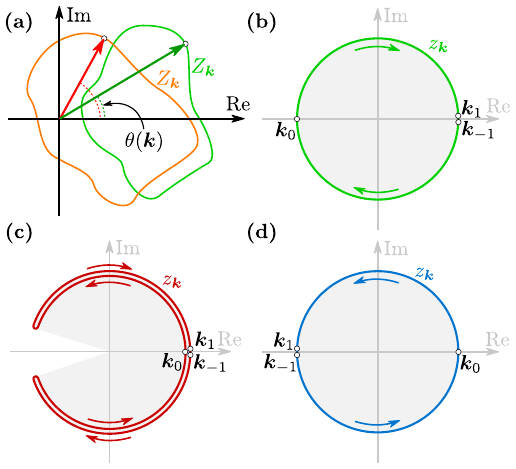}
\caption{
(a) Graphical interpretation of the winding number $w$, given by Eq.~(\ref{eq.wind_def}).
$z_{\bm k} = Z_{\bm k} / | Z_{\bm k} |$ corresponds to a projection of some closed contour given by $Z_{\bm k}$ on the unit circle.
In the case of the non trivial topological phase, the contour created by $Z_{\bm k}$ lies on the complex plane and contains the origin. 
Then, periodic changes of ${\bm k}$ lead to a full winding of the phase (red circle).
Contrary to this, in the trivial phase, $\theta({\bm k})$  does not perform a full winding as a function of  ${\bm k}$ (the origin  is outside of the $Z_{\bm k}$ trajectory).
Panels from (b) to (d) show exemplary results for $\mu/t = -2$, $-1.25$, and $-0.5$, respectively, for fixed $h/t = 0.5$, $\Delta/t = 0.2$ and $\lambda/t = 0.15$.
TRIM (${\bm k}_{0} = 0$ and ${\bm k}_{\pm1} = \pm \pi/2$) are depicted as white points.
The behavior of the winding number in the non trivial phase is shown in  panels (b) and (d). 
For the trivial phase $z_{\bm k}$ does not describe a closed unit circle.
}
\label{fig.wind}
\end{figure}

\section{Real space Bogoliubov--de~Gennes Hamiltonian}\label{app:bdg}

The real space Bogoliubov--de~Gennes (BdG) equations, can be written in the form $\mathcal{E}_{n} \Psi_{isn} = \mathbb{H}_{is,js'} \Psi_{js'n}$, where $\mathbb{H}_{is,js'}$ is the Hamiltonian in the matrix form:
\begin{eqnarray}
\label{eq.ham_bdg_real} && \mathbb{H}_{is,js'} = \\
\nonumber &&\left(
\begin{array}{cccc}
H_{is,js',\uparrow} & S_{is,js'}^{\uparrow\downarrow} & \Delta_{is,js'} & 0 \\ 
S_{is,js'}^{\downarrow\uparrow} & H_{is,js',\downarrow} & 0 & \Delta_{is,js'} \\ 
\Delta_{is,js'}^{\ast} & 0 & -H_{is,js',\downarrow}^{\ast} & -S_{is,js'}^{\downarrow\uparrow} \\ 
0 & \Delta_{is,js'}^{\ast} & -S_{is,js'}^{\uparrow\downarrow} & -H_{is,js',\uparrow}^{\ast}
\end{array} \right)  ,
\end{eqnarray}
with the eigenvectors
\begin{eqnarray}
\Psi_{isn} &=& \left( u_{isn\uparrow} , u_{isn\downarrow} , v_{isn\downarrow} , v_{isn\uparrow} \right)^{T} \,.
\end{eqnarray}
For the considered model (cf. Sec.~
\ref{sec.model_real}), the matrix block elements, 
are given by $H_{is,js',\sigma} 
= - t \delta_{ij} \delta_{\langle ss'\rangle} - t \delta_{i-1,j} \delta_{\langle s,s'\rangle}
- [ \mu + \sigma ( h + m_{0} ( \delta_{sA} - \delta_{sB} ) ) ] \delta_{ij} \delta_{ss'} $, the superconductivity is denoted by 
$\Delta_{is,js'} = \Delta \delta_{ij} \delta_{ss'}$ and $S_{is,js'}^{\sigma\sigma'} = - i \lambda 
( \sigma_{y} )_{\sigma\sigma'} \left( \delta_{ij} \delta_{\langle ss'\rangle} - \delta_{i-1,j} \delta_{\langle ss'\rangle} \right)$ 
gives the spin-orbit term.

\section{Topological invariants}
\label{app:topo}

The winding number $w$ can be found starting from the standard chiral invariant~\cite{gurarie.11}
\begin{equation}
w=\frac{1}{4\pi \im}\int_{-\frac{\pi}{2}}^{\frac{\pi}{2}}dk\Tr\mathcal{S}\tilde{\mathcal{H}}\partial_k\tilde{\mathcal{H}}^{-1}\,,
\end{equation}
which has the equivalent formulation, found after a small amount of manipulation,
\begin{equation}\label{eq.nu_inv}
w=\frac{1}{2\pi \im}\int_0^\pi dk\Tr\left[\partial_k\mathcal{A}(k)\right]\mathcal{A}^{-1}(k)\,.
\end{equation}
This can be easily calculated numerically to find the chiral invariant. However in the following we will find an analytical formula for the invariant. 
Rewriting this as 
\begin{equation}\label{eq.a_inv}
w=\frac{1}{2\pi \im}\int_0^\pi dk\partial_k\ln\det\left[\mathcal{A}(k)\right]\,,
\end{equation}
we see that the invariant is the winding of $\det\ln\left[\mathcal{A}(k)\right]$ across the Brillouin zone.

\begin{figure}[!b]
  \includegraphics[width=\columnwidth]{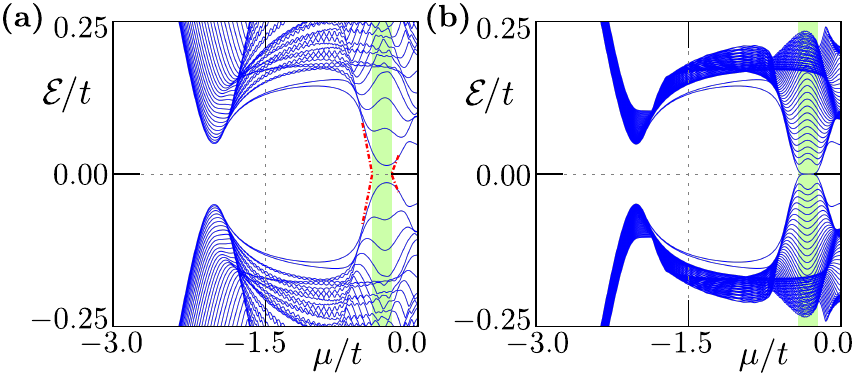}
\caption{
Impact of the chain length on the realization of the non trivial topological phase (marked by the green range of $h$).
The spectrum of the system for $h/t = \pm 0.15$ and $m_{0}/t = 0.2$ as a function of the chemical potential $\mu$.
Results for the chain with 100 (a) and 500 (b) sites.
Only the first 100 eigenvalues around the Fermi level are shown. 
}
\label{fig.eigen_len}
\end{figure}

From the definition of $Z_{\bm k}$ and Eq.~\eqref{eq.a_inv} one can see that the winding number of $z_{\bm k} = Z_{\bm k} / | Z_{\bm k} | = \exp ( i \theta_{\bm k} )$ is equivalently the invariant $w$ and
\begin{eqnarray}
\label{eq.wind_def}
w = \frac{-i}{2\pi} \int_{{\bm k} = -\pi/2}^{{\bm k} = \pi/2} \frac{ dz_{\bm k} }{ z_{\bm k} } = \frac{1}{2\pi} \int_{-\pi/2}^{\pi/2} d{\bm k}\frac{ d\theta_{\bm k} }{ d{\bm k} }\,.
\end{eqnarray}
This clearly takes only integer values (including zero) since $z_{-\pi/2} = z_{\pi/2}$.
The winding number is associated with the number of times that the angle $\theta_{\bm k}$ winds about the origin in the complex plane (see Fig.~\ref{fig.wind}).
This quantity is invariant under smooth perturbation and cannot changed unless $| Z_{\bm k} |$ goes to zero due to gap closing (provided the chiral symmetry is preserved).
The winding number $w$ is the $\mathbb{Z}$ topological index.

\section{Finite size effects}
\label{app:size}

Depending on the length of the localization of the MBS, which in turn depends on the size of the gap, one may need larger system sizes in order to adequately capture the MBS. In Fig.~\ref{fig.eigen_len}, we compare the energy spectrum for system lengths of 100 and 500 sites. In the non trivial topological phase (range of $\mu$ marked by green area), one can see that for the shorter nanowire the MBS do not fully form due to their energy splitting caused by the MBS overlapping in the nanowire [Fig.~\ref{fig.eigen_len}(a)]. However, for a longer nanowire [Fig.~\ref{fig.eigen_len}(b)] this is no longer a problem and there are well formed zero energy states.

\bibliography{biblio}

\end{document}